\pdfoutput=1
\documentclass[11pt]{article}

\usepackage[preprint]{acl}

\usepackage{times}
\usepackage{latexsym}

\usepackage[T1]{fontenc}
\usepackage[utf8]{inputenc}
\usepackage{microtype}
\usepackage{inconsolata}

\usepackage{graphicx}
\usepackage{amsmath}
\usepackage{amssymb}
\usepackage{booktabs}
\usepackage{algorithm}
\usepackage{algorithmic}
\usepackage{array, multirow}
\usepackage{xcolor}
\usepackage{amssymb}
\usepackage[percent]{overpic}
\usepackage{enumitem}
\usepackage{CJKutf8}
\usepackage{xcolor}
\usepackage{svg}
\usepackage{subfig}
\usepackage{comment}

\usepackage[capitalize]{cleveref}

\crefname{section}{Sec.}{Secs.}
\Crefname{section}{Section}{Sections}
\Crefname{table}{Table}{Tables}
\crefname{table}{Tab.}{Tabs.}

\usepackage{xspace} % To handle spacing after macros smartly

\setlist[enumerate]{
  label={\upshape(\roman*)},
  labelwidth=*,
  nosep
}

\newcommand{\eg}{\textit{e.g.,} \xspace}
\newcommand{\ie}{\textit{i.e.,}\xspace}

\title{Cultural Value Differences of LLMs: Prompt, Language, and Model Size}

\author{Qishuai Zhong\quad Yike Yun   \quad Aixin Sun\\
 Nanyang Technological University, Singapore}
\date{May 2024}

\begin{document}
\maketitle

%%%%%%%%% ABSTRACT
\begin{abstract} 
Our study aims to identify behavior patterns in cultural values exhibited by large language models (LLMs). The studied variants include question ordering, prompting language, and model size. Our experiments reveal that each tested LLM can efficiently behave with different cultural values. More interestingly: (i) LLMs exhibit relatively consistent cultural values when presented with prompts in a single language. (ii) The prompting language \eg Chinese or English, can influence the expression of cultural values. The same question can elicit divergent cultural values when the same LLM is queried in a different language. (iii) Differences in sizes of the same model (\eg Llama2-7B vs 13B vs 70B) have a more significant impact on their demonstrated cultural values than model differences (\eg Llama2 vs Mixtral). Our experiments reveal that query language and model size of LLM are the main factors resulting in cultural value differences.

\end{abstract}

%======================================
\section{Introduction}
\label{sec:intro}
%======================================

Since GPT-3~\cite{brown2020language}, Large Language Models (LLMs), capable of generating human-like text based on instructions, have garnered significant attention from both academia and industry. Numerous benchmarks and datasets have been created and employed to assess LLMs' capability in generating human-like text across various tasks like question-answering, chatbot, and summarization~\cite{Clark2018ThinkYH,zellers2019hellaswag,hendryckstest2021}. The open-source leaderboard~\cite{Open-LLM-Leaderboard-Report-2023} allows researchers and engineers to directly compare language models across various dimensions, spanning from commonsense reasoning to advanced question answering, showcasing their respective abilities. However, focusing on information content while ignoring language's social factors is currently a limitation of natural language processing (NLP)~\cite{hovy-yang-2021-importance}. 

Given their capacity to generate human-like text, it is imperative to investigate whether LLMs demonstrate human-like behaviors stemming from the internalized values and cultural insights acquired from large-scale training corpora. As model-generated text gains wider adoption, ethical concerns arise due to the potential influence of cultural biases embedded in the generated text on its users~\cite{kumar-etal-2023-language}. Hence, an emerging research trend involves quantifying the cultural biases within language models and understanding their impact on the models' performance across various tasks. 

The primary methods for assessing values in LLMs typically involve using social science and psychological instruments originally designed for humans~\cite{feng-etal-2023-pretraining,arora2023probing} to assess various cultural aspects quantitatively, or by developing specialized datasets to examine model biases~\cite{parrish2022bbq, huang2023cbbq}.  Many studies on social science instruments primarily evaluate text generated by models in English. However, historical linguistic research, such as the Whorfian hypothesis proposed by Sapir-Whorf, suggests that language structure significantly influences individual perceptions and worldviews~\cite{Kay1984-hk}. Research has shown that cultural accommodation occurs when individuals engage in multilingual contexts, as evidenced by experiments with human subjects~\cite{doi:10.1080/14708470208668081}. Similarly, multilingual language models, pre-trained on text from various languages, can inherit biases and inconsistencies from their training data~\cite{app11073184}. Therefore, assessments using only English-based instruments may not fully capture the breadth of knowledge in multilingual models.

To provide a more comprehensive understanding of LLMs' cultural values, this study investigated patterns of cultural values expressed by different models using three distinct approaches: (i) experimenting with varied prompts in a single language, (ii) using prompts in different languages, and (iii) conducting experiments across different models. The pipeline is visualized in Figure~\ref{fig:overall_pipeline} in Appendix~\ref{appendix:pipeline}. All sets of experiments were designed and implemented using Hofstede's latest Value Survey Module (VSM)~\cite{vsm2013-jj}, a data collection instrument that quantifies cultural values across six dimensions~\cite{TARAS2023101386}. 

A total of 6 LLMs were involved in our experiments, with each model being provided 54 simulated identities to contextualize its response to the VSM questionnaire. Through our investigation, we found that:
(i) LLMs consistently demonstrate similar cultural values within a single language, despite variations in prompt content. However, their responses are affected by alterations in the positioning of options.
(ii) LLMs show notably different cultural values across different languages; and
(iii) Differences in the cultural values expressed by models correlate with variations in text generation proficiency.
For the last finding, although we considered the models' text generation proficiency in our study to conduct our analysis and support our findings, further assessment of the models' generation capability is beyond the scope of our research.

%======================================
\section{Related Work}
\label{sec:related}
%======================================
Several studies have contributed to detecting social and cultural biases displayed by models, as values can be inferred from the expression of biases. Another approach is to incorporate social science models for a direct evaluation of the values inherent in the models. We review both approaches. 

%=====================================
\subsection{Bias Study of Language Model} 
%=====================================

Assessing social and cultural biases in language models is crucial to mitigate associated risks and reveal the values embodied by the models. \citet{pmlr-v139-liang21a} provided a formal comprehension of social biases in language models. The work identified fine-grained local biases and high-level global biases as sources of representational biases and proposed the evaluation metrics for measurement. Subsequently, it introduced the mitigation method. \citet{sheng2021societal} presented the first comprehensive survey on societal biases in language generation in 2021, identifying their negative impact and exploring methods for evaluation and mitigation. The study highlighted the challenge of bias assessment due to the open-domain nature of NLG and the diverse conceptualizations of bias across cultures. Recently, more studies have focused on evaluating bias and values in large language models, with innovative methodologies employed. \citet{cheng2023marked} utilized the concept of markedness, initially linguistic but now a part of social science, to evaluate models' stereotypes unsupervisedly. Meanwhile, \citet{10.1145/3582269.3615599} employed a direct method to assess gender bias in LLMs, revealing models' tendency to reflect imbalances over gender due to training on skewed datasets. In \citet{Ferrara_2023}, bias in generative language models was defined and its sources, such as training data and model specifications, were investigated. However, the study also acknowledged that some biases may persist inevitably due to the inherent nature of language and cultural norms. 

Previous studies have demonstrated diverse techniques for accurately and efficiently identifying biases. However, they have also underscored the challenges in mitigating biases in generated text, as biases can be inherited from human language and culture in training data. This indicates that the exhibited values of models are shaped by the training data, making it impossible to dissociate the influence of training data when trying to understand the patterns of values expressed by models. 

%================================
\subsection{Social Science Models} 
%================================

While studies have investigated language models' social and cultural biases, there's still relatively less systematic exploration of how these models exhibit values under varying circumstances. Quantifying results in this domain is challenging. Consequently, research instruments initially focused on humans have been integrated into understanding language models' values. \citet{feng-etal-2023-pretraining} utilized the political compass test to map the political leaning of language models in a two-dimensional space. Through the experiments conducted, the study demonstrated that pretrained language models are influenced by the political leaning inherent in the training data. Regarding culture measurement, Hofstede's Value Survey Module (VSM)~\cite{vsm2013-jj} and the World Values Survey~\cite{Inglehart2014wvs} were employed by \cite{arora2023probing} to explore cross-cultural values embedded in multilingual masked language models. The evaluation covered 13 languages to probe the models' cultural values across 13 cultures. The findings indicated that pretrained language models captured noticeable differences in values between cultures, albeit with weak correlations to values surveys. 
\citet{kovač2023large} utilized three human psychology questionnaires to assess how models' expression of values changes with varying contexts, such as varying paragraphs and textual formats. They introduced the metaphor ``LLM as a superposition of perspectives" to highlight the context-dependent nature of LLM behavior. 
\citet{shu2023dont} created a dataset covering various persona measurement instruments to evaluate the consistency of LLMs' "personality" across different prompts with minor variations. Their experiments revealed that even minor perturbations notably impacted the models' question-answering performance. Therefore, they argued the current practice of prompting is insufficient to accurately capture model perceptions. The aforementioned articles challenged the practice of using psychological models to reveal personalities by regarding language models as individuals~\cite{bodroza2023personality,pan2023llms}.

Summarized from previous research, it is clear that prompt engineering and training data significantly impact how models express values. However, there is a need for a systematic study to evaluate these factors comprehensively. In our study, we systematically explore the expression of cultural values by models under varying circumstances, including the effects of prompt engineering, language differences, and model capabilities.

%======================================
\section{Measures by VSM}
\label{sec:measures}
%======================================

Similarly to previous studies, we utilize a value survey and additional measurement metrics to evaluate the alignment of cultural values in the LLMs. Value Survey Module (VSM)~\cite{vsm2013-jj} is for measuring cultural values as outlined in Hofstede's Cultural Dimensions Theory~\cite{Gerlach2021-ww}. Despite facing criticism for its psychometric deficiencies~\cite{TARAS2023101386} and simplicity~\cite{Ercan1991-dk}, its value representation has become a cornerstone for a substantial body of research on cross-cultural differences in values~\cite{arora2023probing}. In this study, we utilize the latest version of the survey (VSM 2013) as the foundational assessment.

The value test is structured as a questionnaire with 24 questions to evaluate the interviewees' cultural values. Another six questions intended to gather background information about the interviewees are excluded from our study. The complete questionnaire is in Appendix~\ref{appendix:vsm}. Each question offers respondents five options, labeled with option IDs from 1 to 5. Option IDs also serve as raw scores for each question. The authors of the VSM further developed a scoring system based on each question's raw score, comprising six dimensions for measuring cultural values: Power Distance (\textbf{PDI}), Individualism (\textbf{IDV}), Uncertainty Avoidance (\textbf{UAI}), Masculinity (\textbf{MAS}), Long-term Orientation (\textbf{LTO}), and Indulgence (\textbf{IVR}). Each dimension is calculated using a formula with the raw scores from four survey questions. The complete list of formulas is in Appendix~\ref{appendix:vsm_fomula}. 

All experiments are conducted using prompts derived from the questionnaire. The prompt is delivered in a zero-shot manner, and the LLM is expected to respond in JSON format, specifying the chosen option ID and the rationale behind the selection. We require models to respond with option IDs to mitigate the performance degradation outlined by \citet{zheng2024large}. Prompt samples are depicted in Figure~\ref{fig:prompt_samples} in Appendix~\ref{appendix:pipeline}. In each prompt, we give instructions on the reply format, provide a survey question, and supply a simulated background identity. The simulation provided a target for the model to contextualize the response. Contextual simulation or targeting specific groups of people is a common methodology used by previous studies to guide the generation~\cite{kovač2023large, narayanan-venkit-etal-2023-nationality, ramezani2023knowledge, cheng2023marked}.

%==============================
\subsection{Experiment Set}
\label{sec:exp_group}
%==============================

The experiment conducted in this study consists of multiple \textit{\textbf{experiment set}}s. Each set is defined by a unique combination of three hyper-parameters: (i) the tested LLM, (ii) the prompt language, and (iii) whether options are shuffled. 

Within each experiment set, the language model was presented with a curated collection of simulated identities, each comprising three variables: (i) nationality, (ii) age, and (iii) gender to furnish context for the model's responses to questions. The study encompasses nine nationalities (refer to the full list in Appendix~\ref{appendix:nationality_list}), two genders, and three age groups (25, 35, 45), resulting in a total of 54 identities. These variables align with the VSM survey, encompassing interviewees from various countries, genders, and ages. The chosen nations are globally diverse, representing a range of cultures. To prevent coincidence, each question was queried ten times with different seeds. Consequently, we could collect $10 \times 24 \times 54 = 12960$ responses for each experiment set. During the analysis, we calculate the average of the ten outputs as the final output for a simulated identity, which is used as a single data point (a 24-d vector) in the experiment set.

%=======================================
\subsection{Measures by VSM Raw Scores}
\label{sec:pearson_correlation}
%=======================================
Each set of responses from a simulated identity is represented as a 24-dimensional vector, essential for comparisons within and between experiment sets. To evaluate the strength of relationships between these groups, we calculate the \textit{Pearson correlation coefficients} ($\rho$) and $p$-values among the centroid vectors. The $\rho$ values help determine whether two response vectors exhibit a high correlation, indicating a \textit{shared and similar pattern} in their responses. The $p$-value tests the null hypothesis that no relationship exists between the two compared vectors. If $p < 0.05$, then we reject the null hypothesis and conclude that \textit{there is a significant relationship between the vectors}.

%==================
\subsection{Measures by VSM Scores}
%==================
Using VSM formulas in Appendix~\ref{appendix:vsm_fomula}, we can generate 6-dimensional score vectors (\ie PDI, IDV, UAI, MAS, LTO, and IVR) from the 24-d vectors.

\paragraph{Intra-set Disparity Measurement}
\label{sec:intra_set_method}

In Hofstede's research, VSM scores are analyzed nationally to explore cultural value differences between countries. The scores for the nine nations involved in this study are displayed in Appendix~\ref{appendix:human_results}, where each nation's score represents the average of all responses from its interviewees. Similarly, we calculate the national average for model responses of each experiment set. We then use the standard deviation, denoted as $\sigma_m(v_i)$, to assess the dimensional disparity among nations, where $v_i$ represents the dimension. Similar calculations are performed to compute $\sigma_h(v_i)$ for the human results.

The mean values for each dimension, across all experiment sets and human results, range from $-60$ to $100$, indicating that comparing the disparity between models and human results is reasonable. Appendix~\ref{appendix:avg_vsm_score} provides the complete list of mean values.

We then define the distance among nations observed for humans as 
$D_h$ (see Eq.~\ref{eqn:dh}, where $V$ represents the list of dimensions). This illustrates the variations in cultural values observed among humans. Similarly, the overall disparity among nations observed in each experiment set is denoted as $D_m$. Then, we define the ratio of $D_m$ over $D_h$ as the ``\textbf{Model Cultural Disparity (MCD)}.", shown in Eq.~\ref{eqn:mcd}: 
\begin{align}
    D_h &= \frac{1}{|V|} \sum_{v_i \in V}(\sigma_h(v_i))
    \label{eqn:dh}\\
    D_m &= \frac{1}{|V|} \sum_{v_i \in V}(\sigma_m(v_i))\\
    MCD &= \frac{D_m}{D_h}
    \label{eqn:mcd}
\end{align}
MCD compares the dispersion of cultural values exhibited by models based on simulated nations to that observed among humans in Hofstede's study.

\paragraph{Inter-set Disparity Measurement}
\label{sec:inter_set_method}

The intra-set disparity underscores the impact of contextual information on the models' expression of cultural values. Furthermore, our pipeline uses inter-set disparity to explore how changes in any of the three hyper-parameters—shuffling of options, language, and the tested model—affect the expression of cultural values.

We employ clustering methodologies, \textbf{Davies-Bouldin Index (DBI)}~\cite{4766909} and the \textbf{Silhouette Score ($SS$)}~\cite{ROUSSEEUW198753}, to assess the effectiveness of separation between each pair of experiment sets. Detailed descriptions of the two metrics can be found in Appendix~\ref{appendix:clustering_measurement}. In our study, we pre-define the model responses from each experiment set as a cluster, comprising 54 data points, as detailed in Section~\ref{sec:exp_group}.

Additionally, we have introduced a new measurement method, the \textbf{Silhouette Score with Human Reference ($SS_h$)}, to measure the absolute disparity between pairs of sets, taking human results as the reference point.
    
$SS_h$ is designed based on the Silhouette Score, utilizing nationally aggregated average VSM scores:
\begin{align}
    a_h(n_i)& = \frac{1}{| C_h | -1} \sum_{n_j \in C_h ,i \neq j} d(n_i, n_j) \\
    SS_h &= \frac{1}{2N} \sum  \frac{b(n_i) - a(n_i)}{a_h(n_i)} 
\end{align}
where \(a_h(n_i)\) signifies the mean distance from that nation \(n_i\) to all other nations in the human results. \(a(n_i)\) represents the mean distance from the \(i^{th}\) nation to all other nations within the same experiment set. \(b(n_i)\) denotes the mean distance from the same nation to all nations in another experiment set. Additionally, \(N\) denotes the consistent number of nations involved in this study.
Unlike the previous two metrics, which concentrate solely on the compared set-pair, \(SS_h\) measures the effectiveness of the separation between sets by referencing human results. An $SS_h$ value exceeding one suggests that the separation between the two sets is more pronounced than the disparity observed among humans from various nations.

%============================
\section{Experiment Setting and RQs}
\label{sec:expDesign}
%============================

Given that the value survey is structured as a questionnaire, we have specifically chosen and employed models fine-tuned for chat purposes for this study. A total of six models are evaluated in this study,  including members of the Llama2 family~\cite{touvron2023llama}: Llama2-7b-chat-hf, Llama2-13b-chat-hf, and Llama2-70b-chat-hf; members of the Qwen family~\cite{bai2023qwen}: Qwen-14b-chat and Qwen-72b-chat; and Mixtral-8x7B-Instruct-v0.1~\cite{jiang2024mixtral}, which features a different architecture from the other models. 

All experiments were conducted using Vllm~\cite{kwon2023efficient} with Transformers~\cite{wolf2020huggingfaces} to achieve faster inference. We utilized four Nvidia A6000 cards with CUDA 12.2. We used the default config.json and framework parameters for the models' text generation.

Through experiments, we aim to gain insights into three Research Questions (RQs).

\noindent\textbf{RQ1}: \textit{Can large language models consistently express cultural values when presented with perturbed questions in a single language?} We focus on how responses of the same model vary with changes in contextual information and option order shuffling.

\noindent\textbf{RQ2}: \textit{How does language affect the expression of cultural values in models?} We examine the consistency of cultural values expressed by models when identical questions are posed in different languages.

\noindent\textbf{RQ3}: \textit{What can we infer about models' expression of cultural values when comparing them?} We evaluate whether models from the same family show more consistent cultural values and investigate how differences in text generation capabilities relate to variations in cultural values.

%====================================
\section{Experiment Results}
\label{sec:expResults}
%====================================

Among the 24 questions in the VSM 2013 survey, questions 15 and 18 pertain to the interviewee's recent mental and physical health. Consequently, we assign the most neutral option (option 3) to these two questions. We similarly assign option 3 for any unrecognizable responses from the models.

Initially, we requested responses in Chinese when querying models with Chinese prompts. However, about 7\% of Llama2-7b-chat-hf's responses and 24\% of Llama2-13b-chat-hf's responses were unrecognizable, in contrast to other models which had at least 99\% recognizable responses. As a result, these two models are required to respond in English to Chinese prompts.
 
%=================================
\subsection{Prompt Variants (RQ1)}
%=================================

To evaluate a single model's consistency in expressing cultural values within a single language, we developed prompt variations focusing on two aspects: \textbf{simulated identity} and \textbf{options order}. The former modifies only the context presented to the model, whereas the latter entails further prompt engineering. The impact of simulated identity is assessed within the experiment set, while the effectiveness of options order is evaluated through inter-set methods.

\paragraph{Simulated Identity}
Within each experiment set, the model is queried with 54 simulated identities. VSM raw scores and intra-set measurements are used to examine the impact of simulated identities.

Based on raw scores, each tested model consistently produces results with a similar distribution, irrespective of changes in the context. The correlation coefficients for the average score vectors, grouped by context variables in Table~\ref{tab:single_language_identities_results} in the Appendix, indicate that responses across different identities are highly correlated. 

Similarly, the intra-set measurements based on VSM scores presented in Table~\ref{tab:single_language_vsm_results}, show that the simulated nations assigned to the LLMs exhibit significantly less cultural value diversity compared to the differences observed among human interviewees from those nations.

In summary, \textbf{the evaluated models produce responses with relatively consistent cultural values and show limited sensitivity to changes in the context of the prompts}. The cultural values learned from the training corpus help mitigate the effects of variations in the simulated identities provided in the context.

\begin{table}[!hbt]
\centering
\setlength\tabcolsep{4pt}
\renewcommand{\arraystretch}{1.2}
\begin{tabular}{l|c|c|c}
\hline
\textbf{Models} & { \textbf{$DBI$} $\downarrow$ } & {\textbf{$SS$} $\uparrow$ } & {\textbf{$SS_h$} $\uparrow$ } \\
\hline
{Llama2-7b-chat-hf} & 1.837 & 0.169 & 0.430 \\
{Llama2-13b-chat-hf} & 1.694 & 0.205 & 0.228 \\
{Llama2-70b-chat-hf} & 0.658 & 0.572 & 0.574 \\
{Qwen-14b-chat} & 0.981 & 0.409 & \textbf{1.033} \\
{Qwen-72b-chat} & 0.825 & 0.478 & 0.483 \\
{Mixtral-8x7B} & \textbf{0.542} & \textbf{0.641} & 0.680 \\

\hline
\end{tabular}
\caption{Results of three measurements are listed in the table to quantify the disparity between model responses for the two sets, ``Eng w/o shuffled options" and ``Eng w. shuffled options". Figures showing the greatest distinctness are highlighted in bold in each column. }
\label{tab:distance_between_shuffle_no_shuffle}

\end{table}

\paragraph{Shuffled Options} As noted by~\citet{zheng2024large}, LLMs are susceptible to selection bias, primarily due to token bias and, to a lesser extent, position bias, both of which originate from the training data. Accordingly, our experiment maintains the original option IDs and their corresponding text, only altering their positions to minimize token bias. 

We evaluate the consistency of the model's cultural values despite selection bias by analyzing changes in the distribution of raw scores and measuring the inter-set disparity between the "Eng" and "Eng w. Shuffle" experiment sets for each model. 

The Centroid vector of each experiment set represents the distribution of the set. The correlation coefficient and $p$-value are computed between the centroids, with comprehensive results presented in Table~\ref{tab:pearson_correlation_shuffle_no_shuffle} in Appendix. These results indicate that most models maintain highly correlated score distributions after option shuffling. However, the overall correlation scores are noticeably lower than those calculated for simulated identities.

The inter-set disparity measurement results, as shown in Table~\ref{tab:distance_between_shuffle_no_shuffle}, display the effect of shuffling to models from the aspect of VSM score. The results of $DBI$ and $SS$ suggest that the experiment sets, pre-divided based on shuffling, are clustered but not distinctly separated from each other. When evaluating the disparity between sets using $SS_h$, we find that most models exhibit a noticeable absolute shift in cultural values between the sets, which does not correspond to the significant differences observed among humans from diverse nations. 

\textbf{The experiment results show that models remain vulnerable to selection bias}, consistent with the findings reported in~\cite{zheng2024large}. Unlike human behavior, models fail to maintain consistent cultural values in the face of textual ambiguities. 

\begin{figure*}
    \setlength\tabcolsep{2pt}
    \centering
    \begin{tabular}{ccc}
        \subfloat[Llama2-7b-chat-hf]{\includegraphics[width=0.31\textwidth]{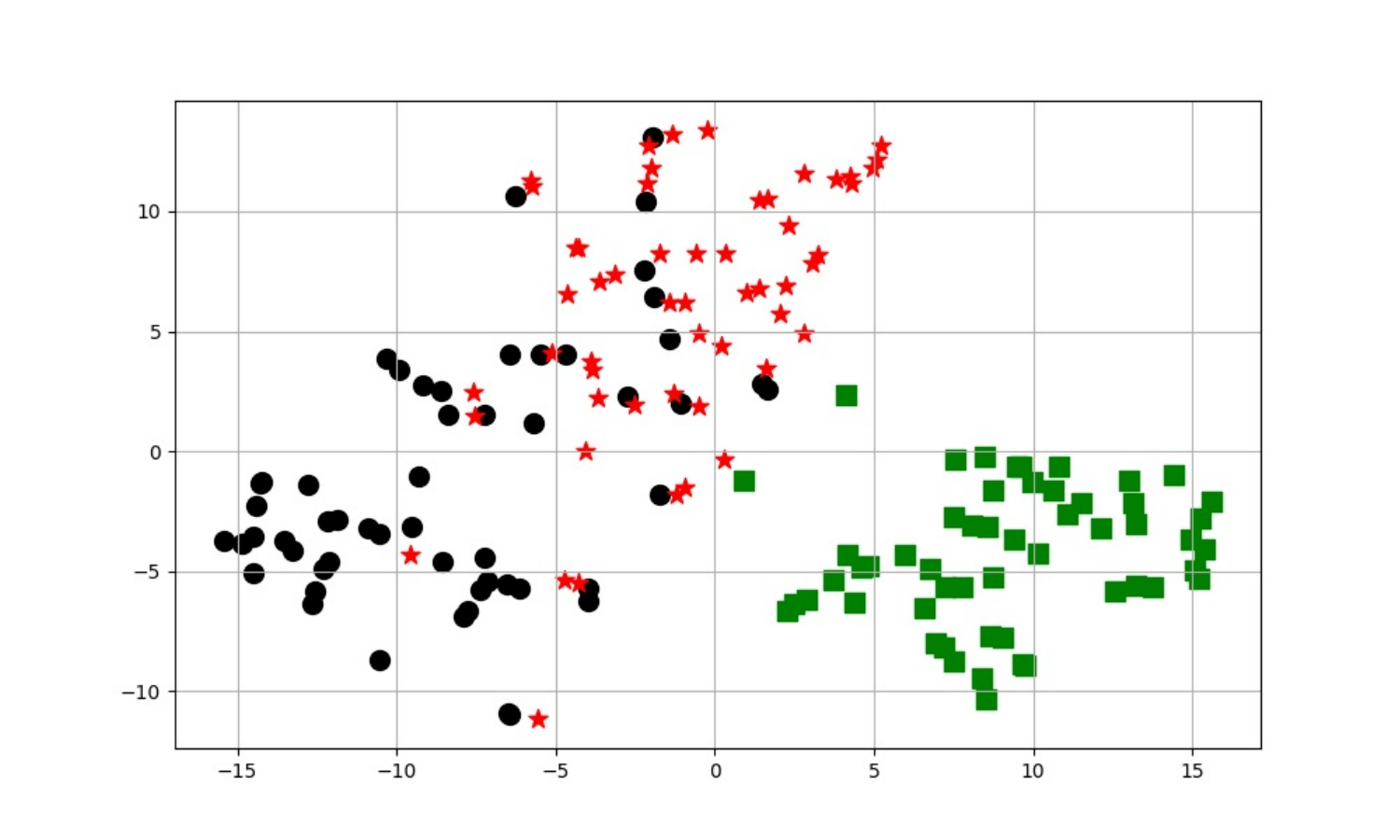}} &
        \subfloat[Llama2-13b-chat-hf]{\includegraphics[width=0.31\textwidth]{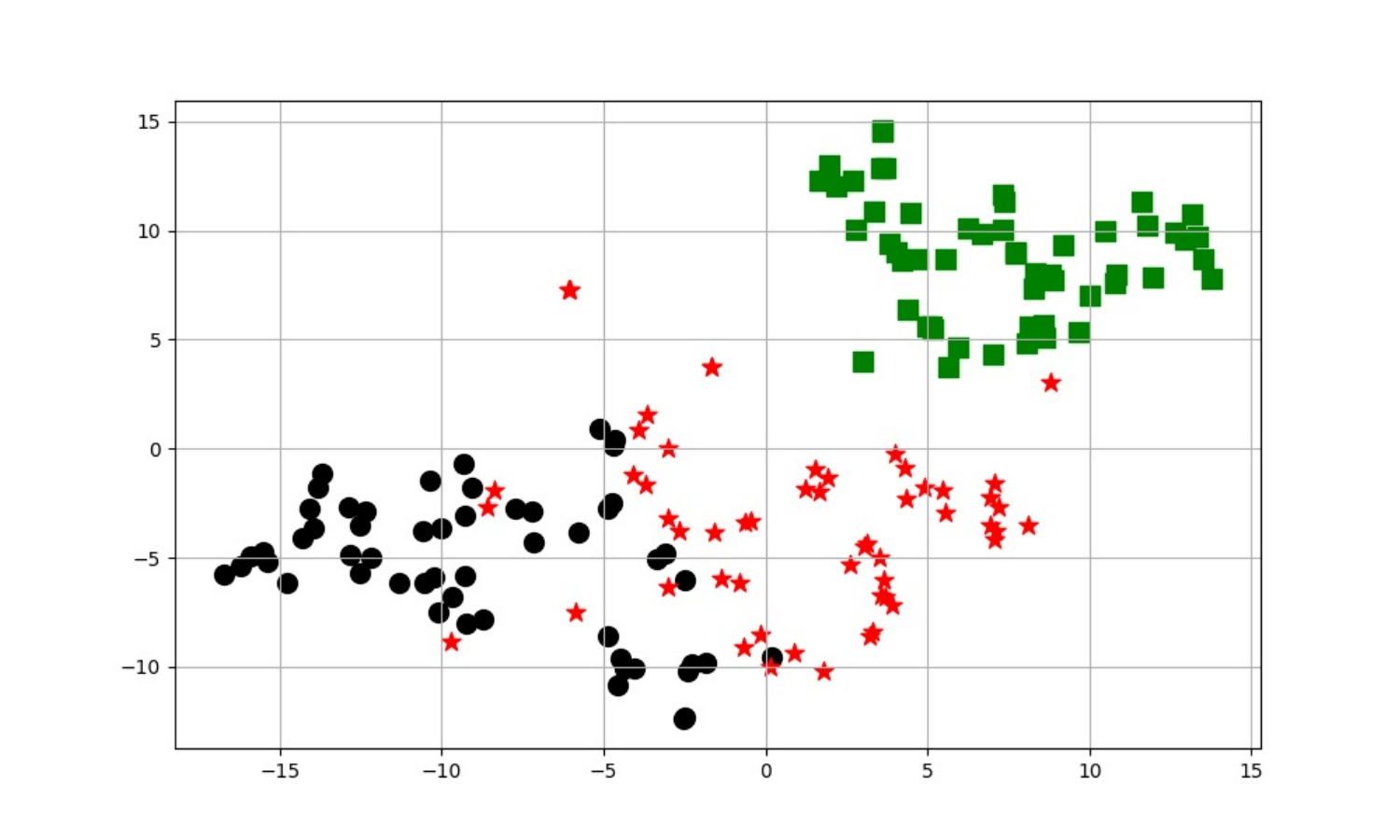}} &
        \subfloat[Llama2-70b-chat-hf]{\includegraphics[width=0.31\textwidth]{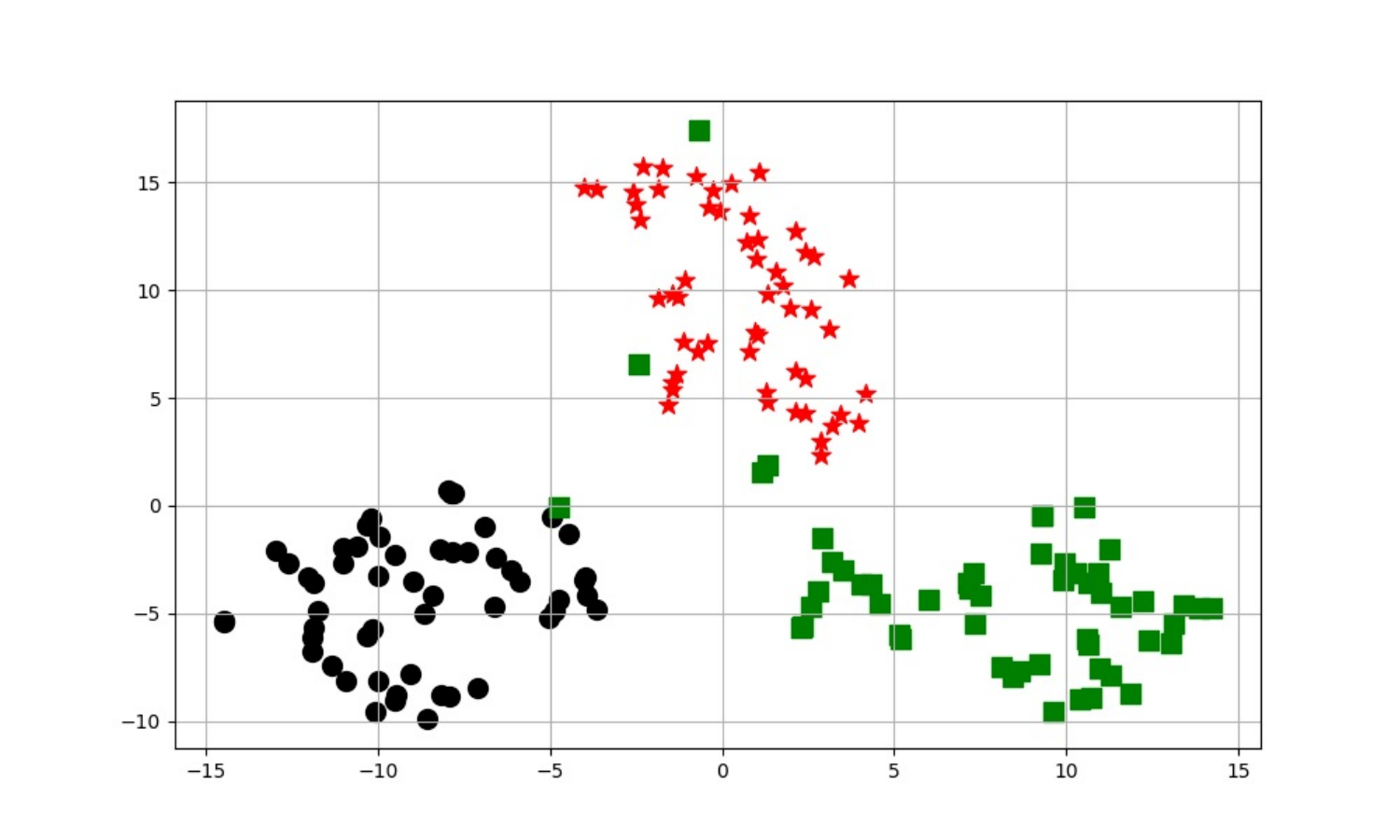}}  \\
        \subfloat[Qwen-14b-chat]{\includegraphics[width=0.31\textwidth]{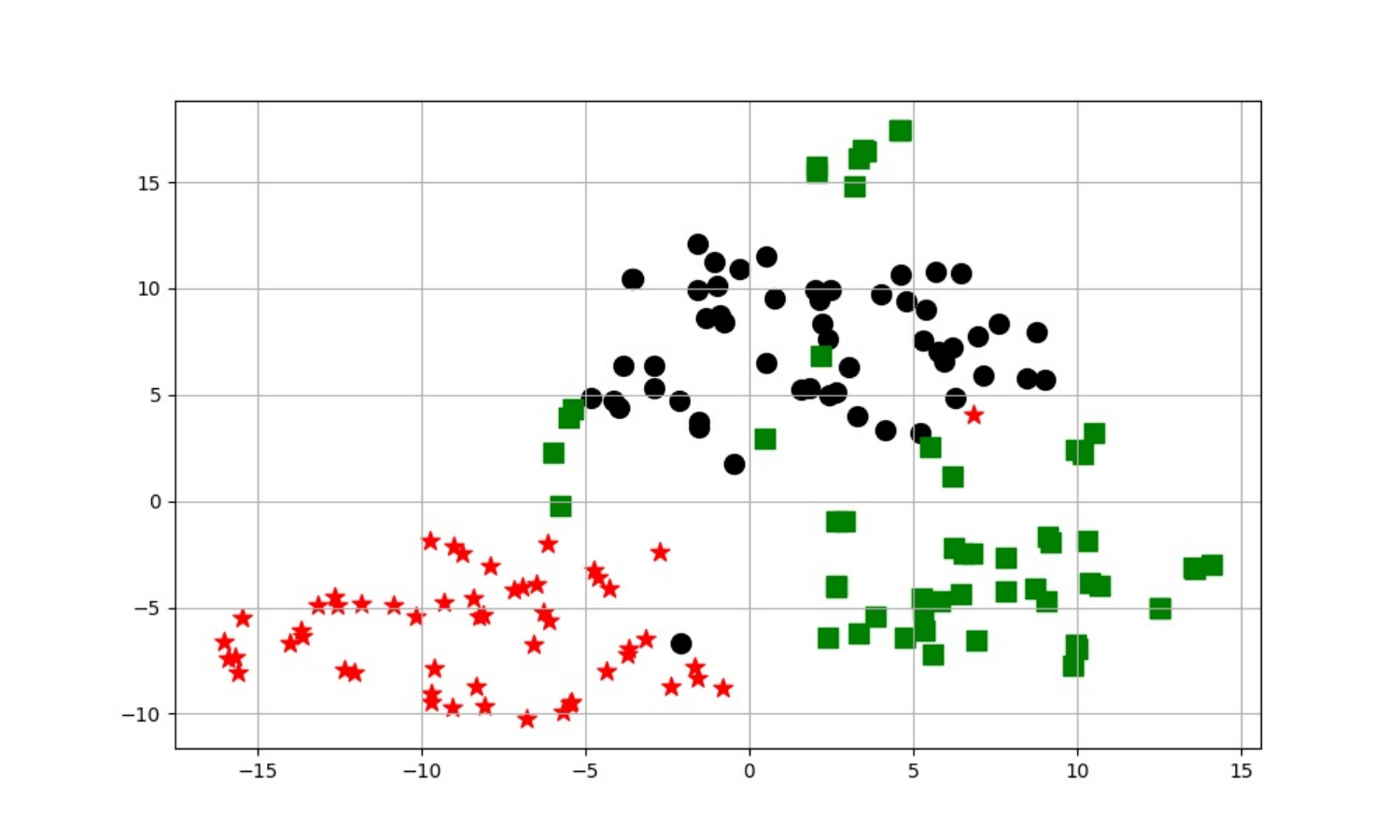}} &
        \subfloat[Qwen-72b-chat]{\includegraphics[width=0.31\textwidth]{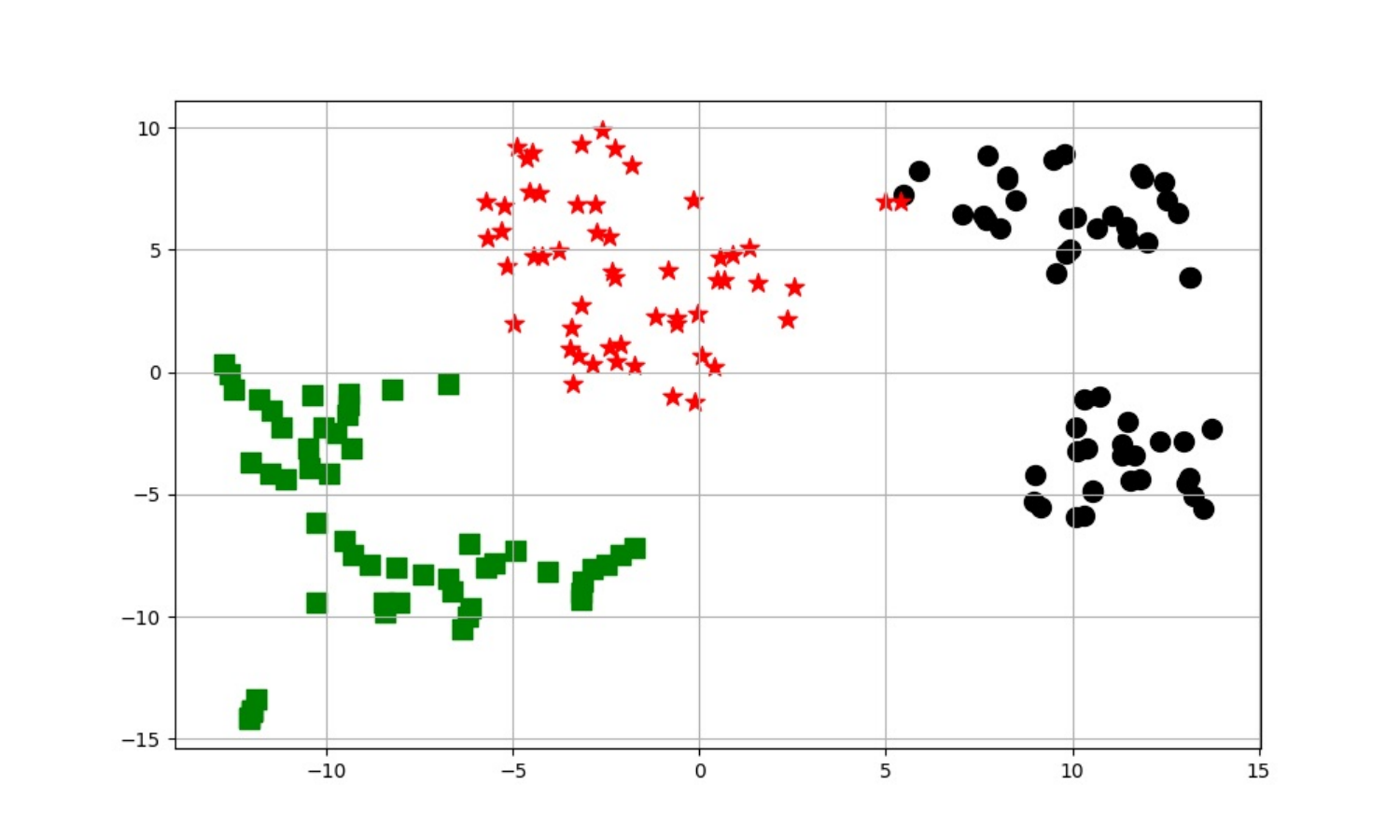}} &
        \subfloat[Mixtral-8x7B]{\includegraphics[width=0.31\textwidth]{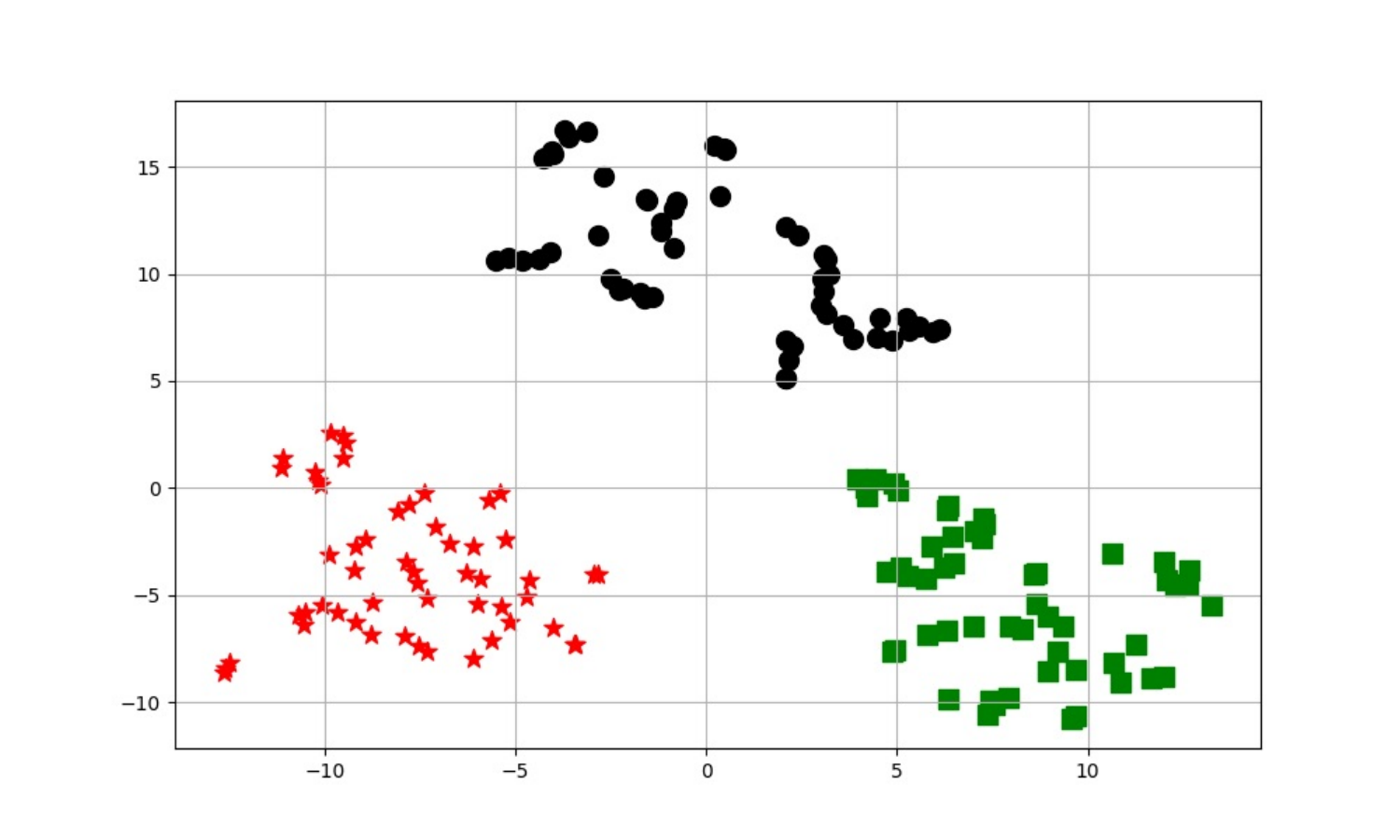}} \\
        \multicolumn{3}{c}{\includegraphics[width=0.5\textwidth]{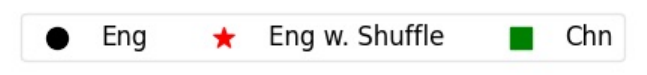}}
    \end{tabular}
    \caption{The 6-d VSM scores for different experiment sets for each model are visualized using the t-SNE technique~\cite{JMLR:v9:vandermaaten08a} to facilitate direct comparisons. Results from English queries (denoted as "Eng") are displayed with black circles; results from English with Shuffled Options (denoted as "Eng w. Shuffle") are shown with pink stars; and results from Chinese (denoted as "Chn") are represented by green squares.}
    \label{fig:t-SNE}
\end{figure*}

Model response distributions across different experiment sets are visualized using t-SNE in Figure~\ref{fig:t-SNE}~\cite{JMLR:v9:vandermaaten08a}. The visualization also indicates that most models demonstrate less or comparable separation effectiveness between sets divided by ``Shuffling of options" compared to those split by ``Language".

%========================================
\subsection{Language Variants (RQ2)}
\label{sec:rq2}
%========================================

\begin{table}[!hbt]
\centering
\setlength\tabcolsep{4pt}
\renewcommand{\arraystretch}{1.2}
\begin{tabular}{l|c|c|c}
\hline
\textbf{Models} & { \textbf{$DBI$} $\downarrow$ } & {\textbf{$SS$} $\uparrow$ } & {\textbf{$SS_h$} $\uparrow$ } \\
\hline
{Llama2-7b-chat-hf} & 0.962 & 0.423 & \textbf{1.357} \\
{Llama2-13b-chat-hf} & 0.720 & 0.533 & 0.581 \\
{Llama2-70b-chat-hf} & 0.799 & 0.499 & 0.707 \\
{Qwen-14b-chat} & 1.846 & 0.215 & 0.622 \\
{Qwen-72b-chat} & \textbf{0.529} & \textbf{0.646} & 0.961 \\
{Mixtral-8x7B} & 0.651 & 0.581 & 0.660 \\

\hline
\end{tabular}
\caption{Results of three measurements are listed in the table to quantify the disparity between model responses for the two sets, ``Eng w/o shuffled options" and ``Chn w/o shuffled options". Figures showing the greatest distinctness are highlighted in bold in each column. }
\label{tab:distance_between_languages}

\end{table}

In addition to varying prompts within the same language, we conduct experiments to evaluate each model's behavior when prompted in English and Chinese. For the Chinese queries, we carefully crafted prompts using the Chinese version of the VSM 2013 questionnaires. Contextual information of the simulated identities is manually translated. 

Correlation coefficients and $p$-values of this group of comparisons are displayed in Table~\ref{tab:pearson_correlation_language} in Appendix~\ref{appendix:RQ2_results}. One model exhibits a $p$-value exceeding the threshold, indicating no significant relationship between its outputs for English and Chinese questions.
Although the $p$-values for other models remain below the threshold, the overall correlation coefficient is lower than that observed with prompt variants. This suggests that language impacts the models' choice of options more significantly than the shuffling of option order.

In addition to the raw scores, the inter-set disparity measurement results based on VSM scores are detailed in Table~\ref{tab:distance_between_languages}, with a comprehensive analysis of values provided in Appendix~\ref{appendix:RQ2_results}.
Based on the results of $DBI$ and $SS$, we find no significant differences between comparisons based on language and ``shuffling". However, the $SS_h$ results suggest that when queried with the same questions in a different language, the model is expected to exhibit cultural values with a variability of at least 50\%, akin to that of an individual from another country. Language differences can result in a more distinctive separation in expressing cultural values. The t-SNE figures in Figure~\ref{fig:t-SNE} also clearly illustrate that most models express cultural values more variably when queried in different languages.

The discrepancies between the initial measurements and $SS_h$ stem from differences in their formulas. $DBI$ and $SS$, focus on the ratio between inter-set distance and intra-set disparity. These measurements may not provide robust values if model responses are sparse. In contrast, the $SS_h$ formula considers inter-set distance with human disparity (a constant), providing an absolute measure of inter-set disparity.

Summarizing the findings, we observe that language significantly influences the models' responses and the cultural values expressed by those responses. This observation aligns with research findings~\cite{3d632746-3982-36aa-a022-195deb79974f} that suggest values are commonly conveyed through language. 
We argue that the diverse cultural values expressed by the model in various languages are acquired from the distinct training corpora of those languages, similar to other types of knowledge transferred from training corpora to the language model~\cite{lin2019open, krishna2023downstream}.

\begin{figure*}[t]
\centering
\centering
    \begin{tabular}{cc}
        \subfloat[$SS_h$ among Models with English Questions]{\includegraphics[width=0.4\textwidth]{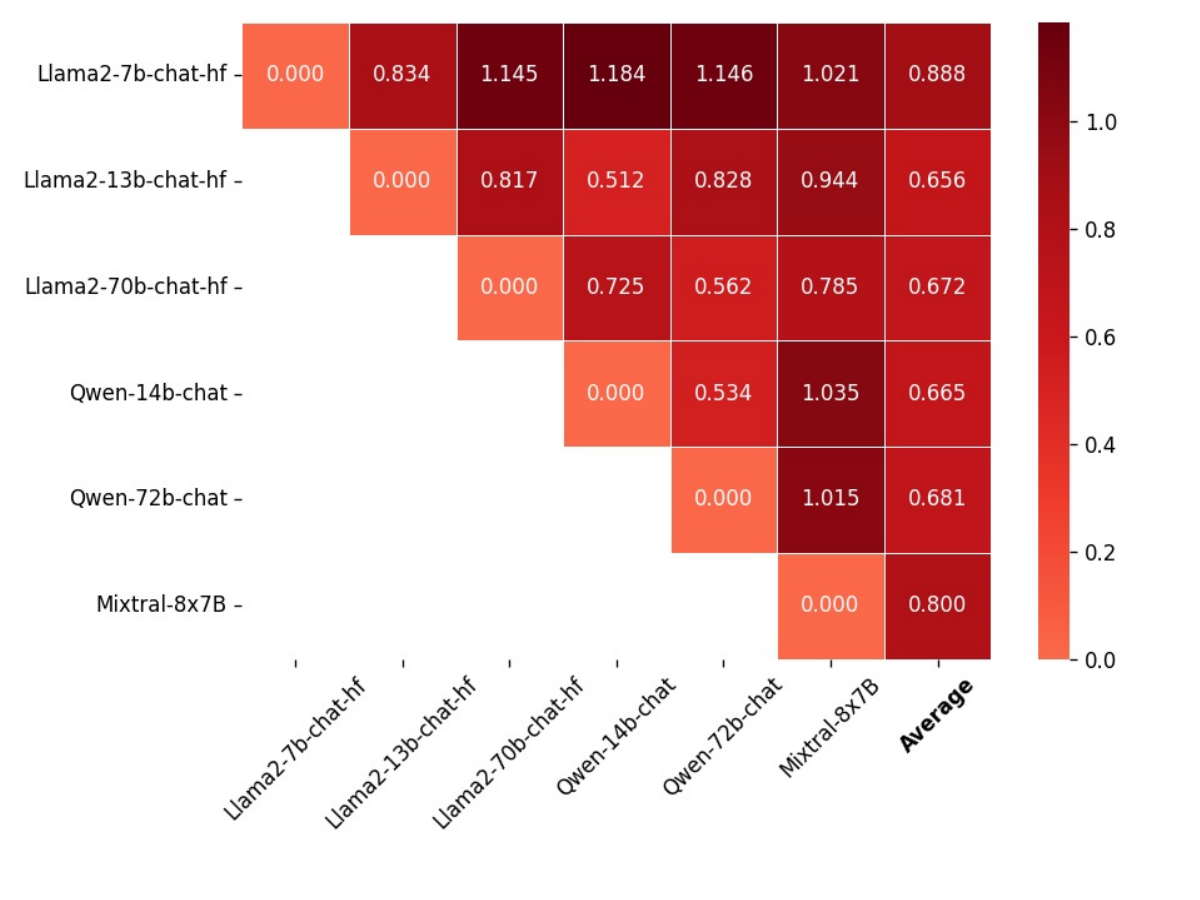}} &
        \subfloat[MMLU Distance among Models]{\includegraphics[width=0.4\textwidth]{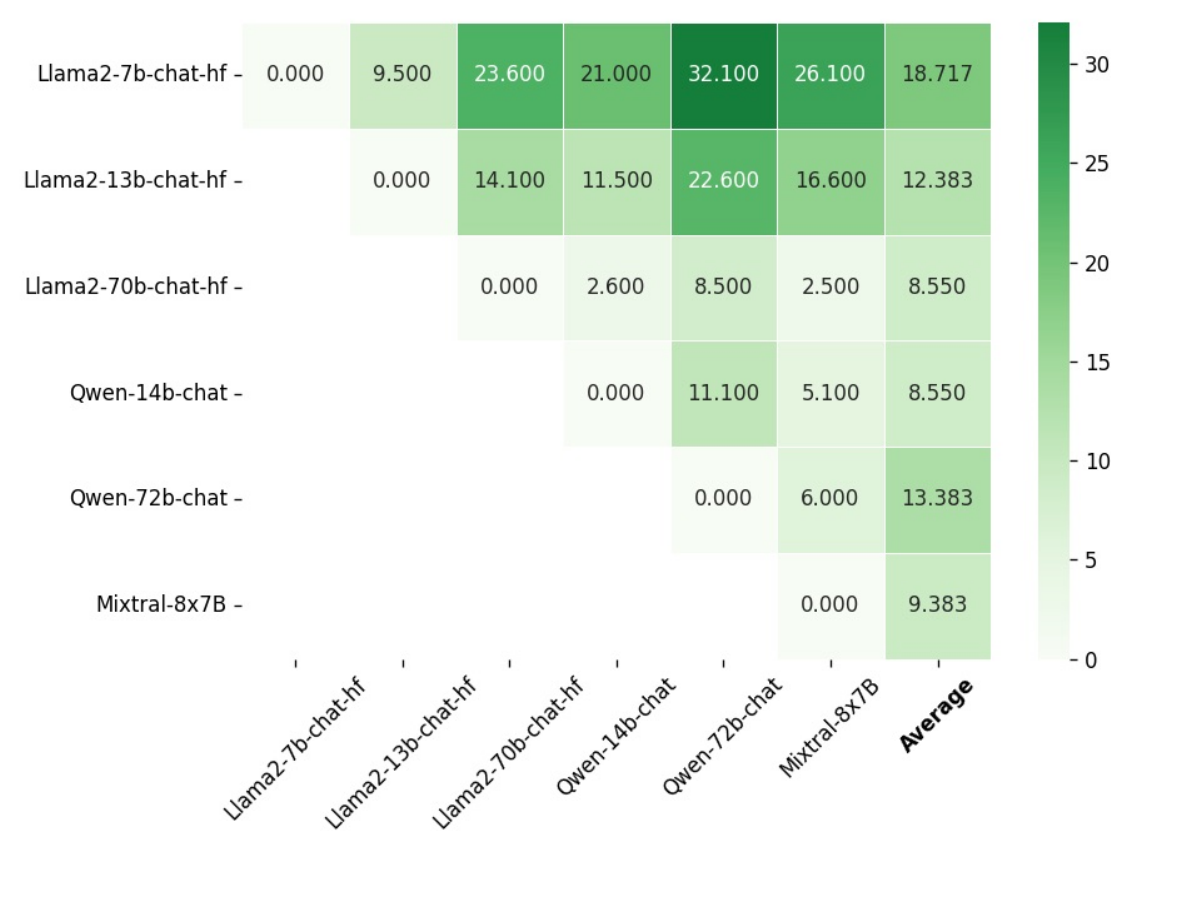}}\\
        \subfloat[$SS_h$ among Models with Chinese Questions]{\includegraphics[width=0.4\textwidth]{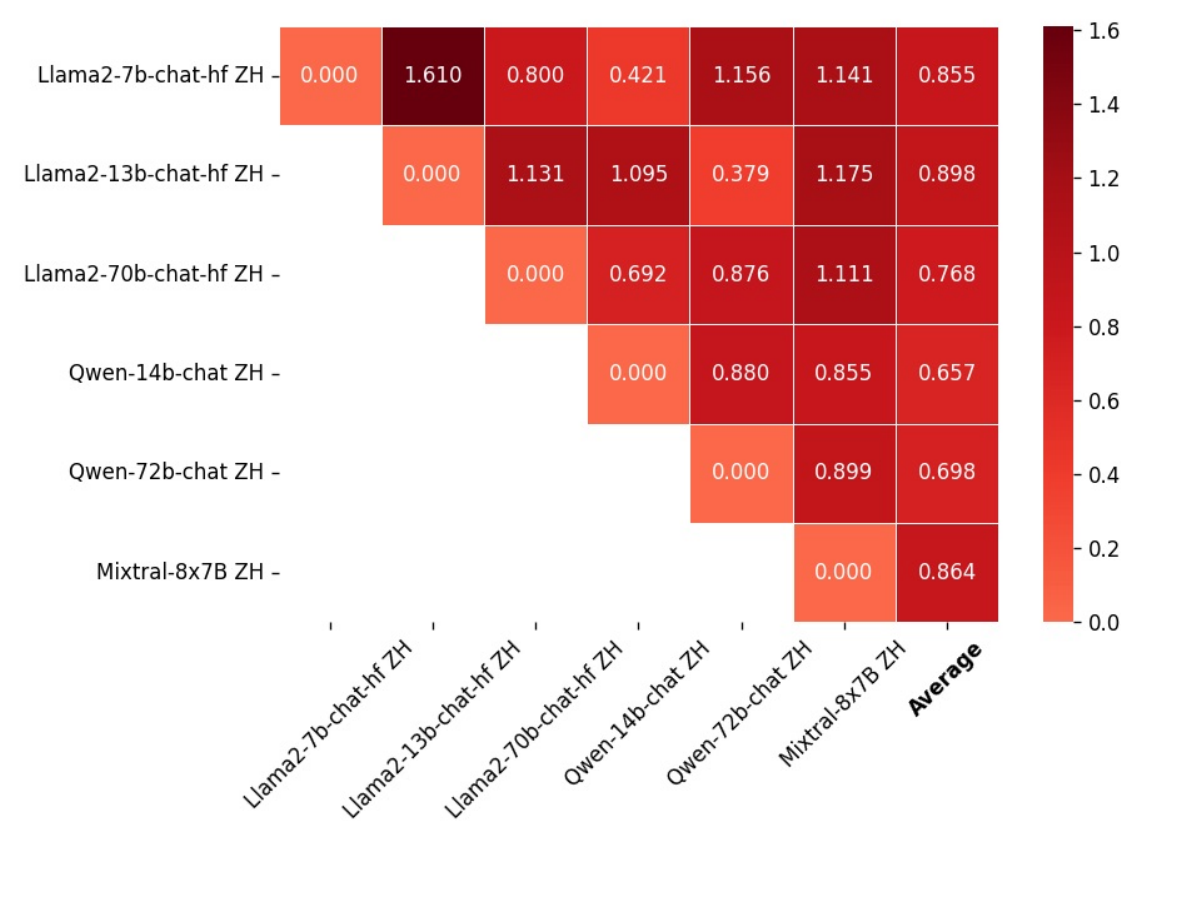}} &
        \subfloat[$SS_h$ among Models cross Languages]{\includegraphics[width=0.4\textwidth]{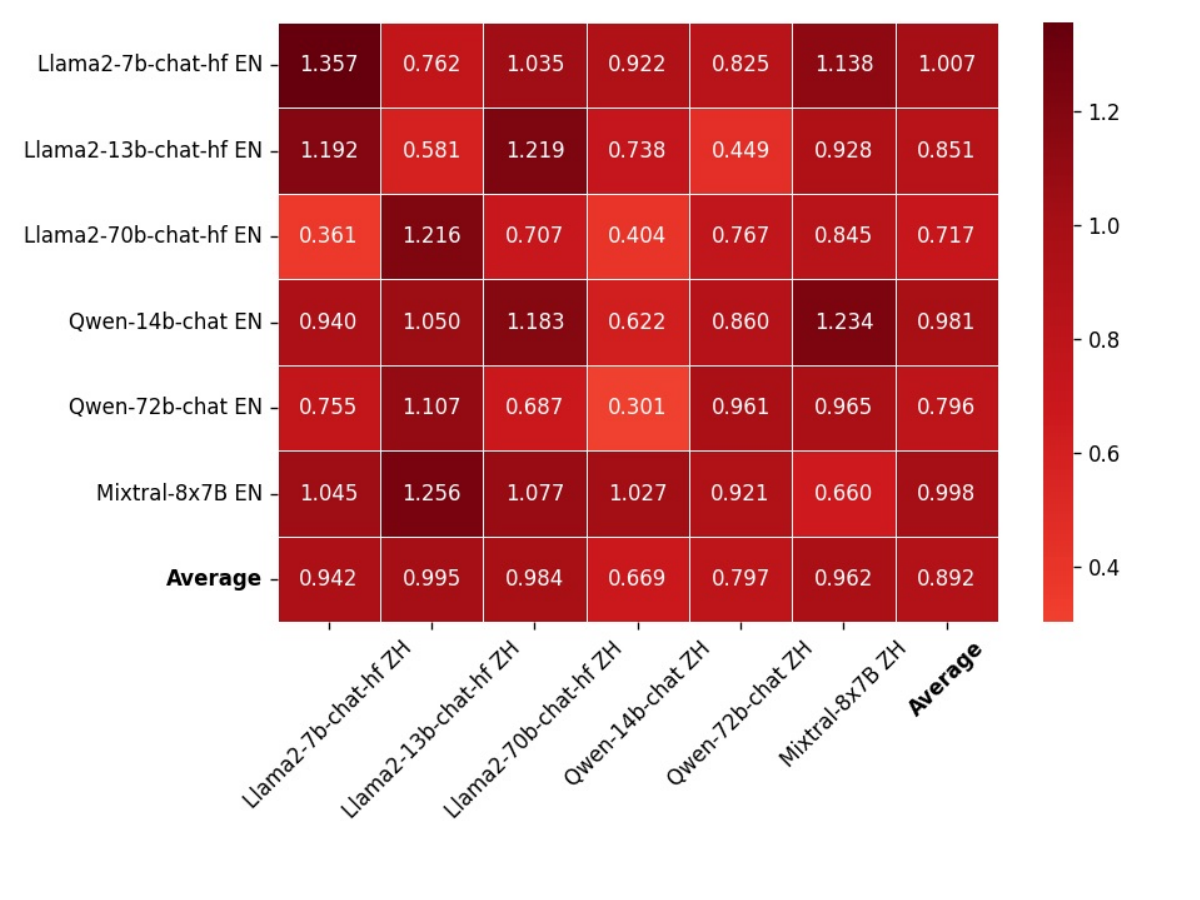}} \\
    \end{tabular}
%TODO: update the caption
\caption{The three red heatmaps display the $SS_h$ values among models, with darker colors highlighting greater disparities. The green heatmap displays the differences in MMLU scores among models, corresponding to the disparities observed in the adjacent red heatmap.}
\label{fig:heatmaps_vsm_scores}
\end{figure*}

%=================================
\subsection{Models Comparison (RQ3)}
%=================================

We now analyze the patterns of cultural values expressed by different models based on their inter-set disparity. This analysis encompasses three types of comparisons: (i) among models queried solely in English (without ``shuffling"), (ii) among models queried solely in Chinese, and (iii) cross-language comparisons. All comparisons utilize $SS_h$ values. We represent all three comparison subsets with heatmap charts, as shown in Figure~\ref{fig:heatmaps_vsm_scores}.

Observations from Heatmaps (a) and (c) in Figure~\ref{fig:heatmaps_vsm_scores} reveal that models from the same family do not necessarily exhibit closer cultural value alignment. Additionally, all Llama2 models, irrespective of size, are trained using the same datasets for the same duration \cite{touvron2023llama}. The Qwen technical report \cite{bai2023qwen} also indicates that identical datasets and hyperparameters are applied across various model sizes during pretraining and fine-tuning stages (SRF and RLHF).

Based on the findings: (i) models from the same family do not guarantee consistency in expressing cultural values; (ii) models with the same background receive uniform training; and (iii) larger models within the same family demonstrate better text-generation performance. We can deduce that \textbf{variations in cultural values among models of the same family are linked to differences in their text-generation capabilities instead of training data}. Larger models in the same family are guaranteed to handle complex patterns, understand context more effectively, and generalize better to unseen data. As a result, they are more adept at comprehending questions posed in value tests and generating more appropriate responses compared to smaller models.

We further link our findings with the evaluation results of generation. A common evaluation all six models have undergone is the MMLU (Massive Multitask Language Understanding) test~\cite{hendrycks2021measuring}. Differences in MMLU scores among models are displayed in Figure~\ref{fig:heatmaps_vsm_scores}, Heatmap (b). A large $SS_h$ value between two models often corresponds with a significant gap in MMLU scores. However, the reverse is not necessarily true: a small gap in MMLU scores does not guarantee a small $SS_h$ value between models.

Additionally, in the heatmap (d) of Figure~\ref{fig:heatmaps_vsm_scores}, the overall disparities between models across languages are significantly larger than those observed within a single language. The marked inter-set disparities noted in cross-language comparisons indicate that language variations can cause substantial differences in cultural values among models.

Our hypothesis that differences in cultural values correlate with variations in model capabilities is based on observations. Developing a testing mechanism that simultaneously evaluates text quality, the expression of cultural values, and their alignment is part of future work. This approach will enhance our understanding of how language model performance impacts the expression of cultural values.

%====================================
\section{Conclusion}
%====================================

In this study, we developed an investigative pipeline to assess the behavior of large language models concerning expressions of cultural values. Our results show that (i) Cultural values tend to remain relatively consistent across variations in prompts, especially when changes are limited to content alone. (ii) LLMs exhibit significantly divergent cultural values across different languages, and (iii) The difference in cultural values among models is relevant to variations in the models' overall proficiency in text generation. Furthermore, upon comparing the results illustrating the second and third findings, we find that language variants can lead to greater disparities in cultural values. Language emerges as the most significant factor influencing the cultural values exhibited by the models.

%=======================================
\section{Limitations}
%=======================================
This study has a few limitations that require further investigation in future research. 
(i) We limited our exploration of cultural values expressed by models to the 24 questions of the VSM 2013 survey, which has been criticized for its simplicity. Therefore, future research should consider incorporating additional cultural value surveys to investigate the models' behavior further.
(ii) This study evaluated and assessed only six models. To further validate the findings regarding the models' expression of cultural values and their performance differences, additional models should be explored and included in future studies.
(iii) In our experiments, models are prompted within a narrowly defined context to generate responses in a zero-shot manner, conditioned solely on the provided context. Future studies should extend beyond direct prompts, exploring how models express cultural values when supplied with extensive past experiences and acting as believable agents \cite{park2023generative}.
(iv) A new evaluation pipeline or mechanism needs to be designed to assess and quantify the relationship between specific cultural value patterns and the generated text's quality. This would build upon the current finding that variations in text quality result in different cultural values.
(v) Although we have observed variations in the cultural values of large language models when the same questions are asked in different languages, we have not thoroughly analyzed user preferences concerning these differences. Future research should develop a systematic approach to assess how language-induced disparities in cultural values impact users and to formulate strategies to mitigate any negative effects.

\section{Ethical Consideration}
All experiments described in this study rely on data from the widely recognized Value Survey Module (VSM) 2013~\cite{vsm2013-jj} and utilize open-source language models. While our analysis includes human subject data, it is important to note that this data is derived from the well-established findings of the VSM 2013 study. Additionally, although our research examines the responses of various large language models to assess cultural values, we explicitly avoid ranking these models to maintain objectivity and ethical integrity.

%%%%%%%%% REFERENCES

%\bibliography{custom}
\bibliography{culture}

\begin{thebibliography}{44}
\providecommand{\natexlab}[1]{#1}

\bibitem[{Arora et~al.(2023)Arora, Kaffee, and Augenstein}]{arora2023probing}
Arnav Arora, Lucie-Aimée Kaffee, and Isabelle Augenstein. 2023.
\newblock \href {https://arxiv.org/abs/2203.13722} {Probing pre-trained
  language models for cross-cultural differences in values}.
\newblock \emph{Preprint}, arXiv:2203.13722.

\bibitem[{Bai et~al.(2023)Bai, Bai, Chu, Cui, Dang, Deng, Fan, Ge, Han, Huang,
  Hui, Ji, Li, Lin, Lin, Liu, Liu, Lu, Lu, Ma, Men, Ren, Ren, Tan, Tan, Tu,
  Wang, Wang, Wang, Wu, Xu, Xu, Yang, Yang, Yang, Yang, Yao, Yu, Yuan, Yuan,
  Zhang, Zhang, Zhang, Zhang, Zhou, Zhou, Zhou, and Zhu}]{bai2023qwen}
Jinze Bai, Shuai Bai, Yunfei Chu, Zeyu Cui, Kai Dang, Xiaodong Deng, Yang Fan,
  Wenbin Ge, Yu~Han, Fei Huang, Binyuan Hui, Luo Ji, Mei Li, Junyang Lin, Runji
  Lin, Dayiheng Liu, Gao Liu, Chengqiang Lu, Keming Lu, Jianxin Ma, Rui Men,
  Xingzhang Ren, Xuancheng Ren, Chuanqi Tan, Sinan Tan, Jianhong Tu, Peng Wang,
  Shijie Wang, Wei Wang, Shengguang Wu, Benfeng Xu, Jin Xu, An~Yang, Hao Yang,
  Jian Yang, Shusheng Yang, Yang Yao, Bowen Yu, Hongyi Yuan, Zheng Yuan,
  Jianwei Zhang, Xingxuan Zhang, Yichang Zhang, Zhenru Zhang, Chang Zhou,
  Jingren Zhou, Xiaohuan Zhou, and Tianhang Zhu. 2023.
\newblock \href {https://arxiv.org/abs/2309.16609} {Qwen technical report}.
\newblock \emph{Preprint}, arXiv:2309.16609.

\bibitem[{Bodroza et~al.(2023)Bodroza, Dinic, and
  Bojic}]{bodroza2023personality}
Bojana Bodroza, Bojana~M. Dinic, and Ljubisa Bojic. 2023.
\newblock \href {https://arxiv.org/abs/2306.04308} {Personality testing of
  gpt-3: Limited temporal reliability, but highlighted social desirability of
  gpt-3's personality instruments results}.
\newblock \emph{Preprint}, arXiv:2306.04308.

\bibitem[{Brown et~al.(2020)Brown, Mann, Ryder, Subbiah, Kaplan, Dhariwal,
  Neelakantan, Shyam, Sastry, Askell, Agarwal, Herbert-Voss, Krueger, Henighan,
  Child, Ramesh, Ziegler, Wu, Winter, Hesse, Chen, Sigler, Litwin, Gray, Chess,
  Clark, Berner, McCandlish, Radford, Sutskever, and
  Amodei}]{brown2020language}
Tom~B. Brown, Benjamin Mann, Nick Ryder, Melanie Subbiah, Jared Kaplan,
  Prafulla Dhariwal, Arvind Neelakantan, Pranav Shyam, Girish Sastry, Amanda
  Askell, Sandhini Agarwal, Ariel Herbert-Voss, Gretchen Krueger, Tom Henighan,
  Rewon Child, Aditya Ramesh, Daniel~M. Ziegler, Jeffrey Wu, Clemens Winter,
  Christopher Hesse, Mark Chen, Eric Sigler, Mateusz Litwin, Scott Gray,
  Benjamin Chess, Jack Clark, Christopher Berner, Sam McCandlish, Alec Radford,
  Ilya Sutskever, and Dario Amodei. 2020.
\newblock \href {https://arxiv.org/abs/2005.14165} {Language models are
  few-shot learners}.

\bibitem[{Cheng et~al.(2023)Cheng, Durmus, and Jurafsky}]{cheng2023marked}
Myra Cheng, Esin Durmus, and Dan Jurafsky. 2023.
\newblock \href {https://arxiv.org/abs/2305.18189} {Marked personas: Using
  natural language prompts to measure stereotypes in language models}.
\newblock \emph{Preprint}, arXiv:2305.18189.

\bibitem[{Clark et~al.(2018)Clark, Cowhey, Etzioni, Khot, Sabharwal, Schoenick,
  and Tafjord}]{Clark2018ThinkYH}
Peter Clark, Isaac Cowhey, Oren Etzioni, Tushar Khot, Ashish Sabharwal, Carissa
  Schoenick, and Oyvind Tafjord. 2018.
\newblock \href {https://api.semanticscholar.org/CorpusID:3922816} {Think you
  have solved question answering? try arc, the ai2 reasoning challenge}.
\newblock \emph{ArXiv}, abs/1803.05457.

\bibitem[{Davies and Bouldin(1979)}]{4766909}
David~L. Davies and Donald~W. Bouldin. 1979.
\newblock \href {https://doi.org/10.1109/TPAMI.1979.4766909} {A cluster
  separation measure}.
\newblock \emph{IEEE Transactions on Pattern Analysis and Machine
  Intelligence}, PAMI-1(2):224--227.

\bibitem[{Ercan et~al.(1991)Ercan, Nasif, Al-Daeaj, and
  Ebrahimi}]{Ercan1991-dk}
G~Ercan, Hamad Nasif, Bahman Al-Daeaj, and Mary~S Ebrahimi. 1991.
\newblock Methodological problems in cross-cultural research: An updated
  review.
\newblock \emph{Management International Review}, pages 79--91.

\bibitem[{Feng et~al.(2023)Feng, Park, Liu, and
  Tsvetkov}]{feng-etal-2023-pretraining}
Shangbin Feng, Chan~Young Park, Yuhan Liu, and Yulia Tsvetkov. 2023.
\newblock \href {https://doi.org/10.18653/v1/2023.acl-long.656} {From
  pretraining data to language models to downstream tasks: Tracking the trails
  of political biases leading to unfair {NLP} models}.
\newblock In \emph{Proceedings of the 61st Annual Meeting of the Association
  for Computational Linguistics (Volume 1: Long Papers)}, pages 11737--11762,
  Toronto, Canada. Association for Computational Linguistics.

\bibitem[{Ferrara(2023)}]{Ferrara_2023}
Emilio Ferrara. 2023.
\newblock \href {https://doi.org/10.5210/fm.v28i11.13346} {Should chatgpt be
  biased? challenges and risks of bias in large language models}.
\newblock \emph{First Monday}.

\bibitem[{Garrido-Muñoz  et~al.(2021)Garrido-Muñoz , Montejo-Ráez ,
  Martínez-Santiago , and Ureña-López }]{app11073184}
Ismael Garrido-Muñoz , Arturo Montejo-Ráez , Fernando Martínez-Santiago ,
  and L.~Alfonso Ureña-López . 2021.
\newblock \href {https://doi.org/10.3390/app11073184} {A survey on bias in deep
  nlp}.
\newblock \emph{Applied Sciences}, 11(7).

\bibitem[{Gerlach and Eriksson(2021)}]{Gerlach2021-ww}
Philipp Gerlach and Kimmo Eriksson. 2021.
\newblock Measuring cultural dimensions: External validity and internal
  consistency of hofstede's {VSM} 2013 scales.
\newblock \emph{Front. Psychol.}, 12:662604.

\bibitem[{Harzing and Maznevski(2002)}]{doi:10.1080/14708470208668081}
Anne-Wil Harzing and Martha Maznevski. 2002.
\newblock \href {https://doi.org/10.1080/14708470208668081} {The interaction
  between language and culture: A test of the cultural accommodation hypothesis
  in seven countries}.
\newblock \emph{Language and Intercultural Communication}, 2(2):120--139.

\bibitem[{Hendrycks et~al.(2021{\natexlab{a}})Hendrycks, Burns, Basart, Zou,
  Mazeika, Song, and Steinhardt}]{hendryckstest2021}
Dan Hendrycks, Collin Burns, Steven Basart, Andy Zou, Mantas Mazeika, Dawn
  Song, and Jacob Steinhardt. 2021{\natexlab{a}}.
\newblock Measuring massive multitask language understanding.
\newblock \emph{Proceedings of the International Conference on Learning
  Representations (ICLR)}.

\bibitem[{Hendrycks et~al.(2021{\natexlab{b}})Hendrycks, Burns, Basart, Zou,
  Mazeika, Song, and Steinhardt}]{hendrycks2021measuring}
Dan Hendrycks, Collin Burns, Steven Basart, Andy Zou, Mantas Mazeika, Dawn
  Song, and Jacob Steinhardt. 2021{\natexlab{b}}.
\newblock \href {https://arxiv.org/abs/2009.03300} {Measuring massive multitask
  language understanding}.
\newblock \emph{Preprint}, arXiv:2009.03300.

\bibitem[{Hofstede and Hofstede(2016)}]{vsm2013-jj}
G~Hofstede and G.~J. Hofstede. 2016.
\newblock {VSM} 2013.
\newblock \url{https://geerthofstede.com/research-and-vsm/vsm-2013/}.
\newblock Accessed: 2024-1-11.

\bibitem[{Hovy and Yang(2021)}]{hovy-yang-2021-importance}
Dirk Hovy and Diyi Yang. 2021.
\newblock \href {https://doi.org/10.18653/v1/2021.naacl-main.49} {The
  importance of modeling social factors of language: Theory and practice}.
\newblock In \emph{Proceedings of the 2021 Conference of the North American
  Chapter of the Association for Computational Linguistics: Human Language
  Technologies}, pages 588--602, Online. Association for Computational
  Linguistics.

\bibitem[{Huang and Xiong(2023)}]{huang2023cbbq}
Yufei Huang and Deyi Xiong. 2023.
\newblock \href {https://arxiv.org/abs/2306.16244} {Cbbq: A chinese bias
  benchmark dataset curated with human-ai collaboration for large language
  models}.
\newblock \emph{Preprint}, arXiv:2306.16244.

\bibitem[{Inglehart et~al.(2014)Inglehart, Haerpfer, Moreno, Welzel, Kizilova,
  Diez-Medrano, Lagos, Norris, Ponarin, Puranen, and et~al.}]{Inglehart2014wvs}
R.~Inglehart, C.~Haerpfer, A.~Moreno, C.~Welzel, K.~Kizilova, J.~Diez-Medrano,
  M.~Lagos, P.~Norris, E.~Ponarin, B.~Puranen, and et~al. 2014.
\newblock \href {/brokenurl#www.worldvaluessurvey.org/WVSDocumentationWV6.jsp}
  {World values survey: Round six - country-pooled datafile version}.

\bibitem[{Jiang et~al.(2024)Jiang, Sablayrolles, Roux, Mensch, Savary, Bamford,
  Chaplot, de~las Casas, Hanna, Bressand, Lengyel, Bour, Lample, Lavaud,
  Saulnier, Lachaux, Stock, Subramanian, Yang, Antoniak, Scao, Gervet, Lavril,
  Wang, Lacroix, and Sayed}]{jiang2024mixtral}
Albert~Q. Jiang, Alexandre Sablayrolles, Antoine Roux, Arthur Mensch, Blanche
  Savary, Chris Bamford, Devendra~Singh Chaplot, Diego de~las Casas, Emma~Bou
  Hanna, Florian Bressand, Gianna Lengyel, Guillaume Bour, Guillaume Lample,
  Lélio~Renard Lavaud, Lucile Saulnier, Marie-Anne Lachaux, Pierre Stock,
  Sandeep Subramanian, Sophia Yang, Szymon Antoniak, Teven~Le Scao, Théophile
  Gervet, Thibaut Lavril, Thomas Wang, Timothée Lacroix, and William~El Sayed.
  2024.
\newblock \href {https://arxiv.org/abs/2401.04088} {Mixtral of experts}.
\newblock \emph{Preprint}, arXiv:2401.04088.

\bibitem[{Kay and Kempton(1984)}]{Kay1984-hk}
Paul Kay and Willett Kempton. 1984.
\newblock What is the sapir-whorf hypothesis?
\newblock \emph{Am. Anthropol.}, 86(1):65--79.

\bibitem[{Kotek et~al.(2023)Kotek, Dockum, and Sun}]{10.1145/3582269.3615599}
Hadas Kotek, Rikker Dockum, and David Sun. 2023.
\newblock \href {https://doi.org/10.1145/3582269.3615599} {Gender bias and
  stereotypes in large language models}.
\newblock In \emph{Proceedings of The ACM Collective Intelligence Conference},
  CI '23, page 12–24, New York, NY, USA. Association for Computing Machinery.

\bibitem[{Kovač et~al.(2023)Kovač, Sawayama, Portelas, Colas, Dominey, and
  Oudeyer}]{kovač2023large}
Grgur Kovač, Masataka Sawayama, Rémy Portelas, Cédric Colas, Peter~Ford
  Dominey, and Pierre-Yves Oudeyer. 2023.
\newblock \href {https://arxiv.org/abs/2307.07870} {Large language models as
  superpositions of cultural perspectives}.
\newblock \emph{Preprint}, arXiv:2307.07870.

\bibitem[{Krishna et~al.(2023)Krishna, Garg, Bigham, and
  Lipton}]{krishna2023downstream}
Kundan Krishna, Saurabh Garg, Jeffrey~P. Bigham, and Zachary~C. Lipton. 2023.
\newblock \href {https://arxiv.org/abs/2209.14389} {Downstream datasets make
  surprisingly good pretraining corpora}.
\newblock \emph{Preprint}, arXiv:2209.14389.

\bibitem[{Kumar et~al.(2023)Kumar, Balachandran, Njoo, Anastasopoulos, and
  Tsvetkov}]{kumar-etal-2023-language}
Sachin Kumar, Vidhisha Balachandran, Lucille Njoo, Antonios Anastasopoulos, and
  Yulia Tsvetkov. 2023.
\newblock \href {https://doi.org/10.18653/v1/2023.eacl-main.241} {Language
  generation models can cause harm: So what can we do about it? an actionable
  survey}.
\newblock In \emph{Proceedings of the 17th Conference of the European Chapter
  of the Association for Computational Linguistics}, pages 3299--3321,
  Dubrovnik, Croatia. Association for Computational Linguistics.

\bibitem[{Kwon et~al.(2023)Kwon, Li, Zhuang, Sheng, Zheng, Yu, Gonzalez, Zhang,
  and Stoica}]{kwon2023efficient}
Woosuk Kwon, Zhuohan Li, Siyuan Zhuang, Ying Sheng, Lianmin Zheng, Cody~Hao Yu,
  Joseph~E. Gonzalez, Hao Zhang, and Ion Stoica. 2023.
\newblock \href {https://arxiv.org/abs/2309.06180} {Efficient memory management
  for large language model serving with pagedattention}.
\newblock \emph{Preprint}, arXiv:2309.06180.

\bibitem[{Liang et~al.(2021)Liang, Wu, Morency, and
  Salakhutdinov}]{pmlr-v139-liang21a}
Paul~Pu Liang, Chiyu Wu, Louis-Philippe Morency, and Ruslan Salakhutdinov.
  2021.
\newblock \href {https://proceedings.mlr.press/v139/liang21a.html} {Towards
  understanding and mitigating social biases in language models}.
\newblock In \emph{Proceedings of the 38th International Conference on Machine
  Learning}, volume 139 of \emph{Proceedings of Machine Learning Research},
  pages 6565--6576. PMLR.

\bibitem[{Lin et~al.(2019)Lin, Tan, and Frank}]{lin2019open}
Yongjie Lin, Yi~Chern Tan, and Robert Frank. 2019.
\newblock \href {https://arxiv.org/abs/1906.01698} {Open sesame: Getting inside
  bert's linguistic knowledge}.
\newblock \emph{Preprint}, arXiv:1906.01698.

\bibitem[{Narayanan~Venkit et~al.(2023)Narayanan~Venkit, Gautam, Panchanadikar,
  Huang, and Wilson}]{narayanan-venkit-etal-2023-nationality}
Pranav Narayanan~Venkit, Sanjana Gautam, Ruchi Panchanadikar, Ting-Hao Huang,
  and Shomir Wilson. 2023.
\newblock \href {https://doi.org/10.18653/v1/2023.eacl-main.9} {Nationality
  bias in text generation}.
\newblock In \emph{Proceedings of the 17th Conference of the European Chapter
  of the Association for Computational Linguistics}, pages 116--122, Dubrovnik,
  Croatia. Association for Computational Linguistics.

\bibitem[{Norton(1997)}]{3d632746-3982-36aa-a022-195deb79974f}
Bonny Norton. 1997.
\newblock \href {http://www.jstor.org/stable/3587831} {Language, identity, and
  the ownership of english}.
\newblock \emph{TESOL Quarterly}, 31(3):409--429.

\bibitem[{Pan and Zeng(2023)}]{pan2023llms}
Keyu Pan and Yawen Zeng. 2023.
\newblock \href {https://arxiv.org/abs/2307.16180} {Do llms possess a
  personality? making the mbti test an amazing evaluation for large language
  models}.
\newblock \emph{Preprint}, arXiv:2307.16180.

\bibitem[{Park(2023)}]{Open-LLM-Leaderboard-Report-2023}
Daniel Park. 2023.
\newblock \href {https://github.com/dsdanielpark/Open-LLM-Leaderboard-Report}
  {Open-llm-leaderboard-report}.

\bibitem[{Park et~al.(2023)Park, O'Brien, Cai, Morris, Liang, and
  Bernstein}]{park2023generative}
Joon~Sung Park, Joseph~C. O'Brien, Carrie~J. Cai, Meredith~Ringel Morris, Percy
  Liang, and Michael~S. Bernstein. 2023.
\newblock \href {https://arxiv.org/abs/2304.03442} {Generative agents:
  Interactive simulacra of human behavior}.
\newblock \emph{Preprint}, arXiv:2304.03442.

\bibitem[{Parrish et~al.(2022)Parrish, Chen, Nangia, Padmakumar, Phang,
  Thompson, Htut, and Bowman}]{parrish2022bbq}
Alicia Parrish, Angelica Chen, Nikita Nangia, Vishakh Padmakumar, Jason Phang,
  Jana Thompson, Phu~Mon Htut, and Samuel~R. Bowman. 2022.
\newblock \href {https://arxiv.org/abs/2110.08193} {Bbq: A hand-built bias
  benchmark for question answering}.
\newblock \emph{Preprint}, arXiv:2110.08193.

\bibitem[{Ramezani and Xu(2023)}]{ramezani2023knowledge}
Aida Ramezani and Yang Xu. 2023.
\newblock \href {https://arxiv.org/abs/2306.01857} {Knowledge of cultural moral
  norms in large language models}.
\newblock \emph{Preprint}, arXiv:2306.01857.

\bibitem[{Rousseeuw(1987)}]{ROUSSEEUW198753}
Peter~J. Rousseeuw. 1987.
\newblock \href {https://doi.org/10.1016/0377-0427(87)90125-7} {Silhouettes: A
  graphical aid to the interpretation and validation of cluster analysis}.
\newblock \emph{Journal of Computational and Applied Mathematics}, 20:53--65.

\bibitem[{Sheng et~al.(2021)Sheng, Chang, Natarajan, and
  Peng}]{sheng2021societal}
Emily Sheng, Kai-Wei Chang, Premkumar Natarajan, and Nanyun Peng. 2021.
\newblock \href {https://arxiv.org/abs/2105.04054} {Societal biases in language
  generation: Progress and challenges}.
\newblock \emph{Preprint}, arXiv:2105.04054.

\bibitem[{Shu et~al.(2023)Shu, Zhang, Choi, Dunagan, Card, and
  Jurgens}]{shu2023dont}
Bangzhao Shu, Lechen Zhang, Minje Choi, Lavinia Dunagan, Dallas Card, and David
  Jurgens. 2023.
\newblock \href {https://arxiv.org/abs/2311.09718} {You don't need a
  personality test to know these models are unreliable: Assessing the
  reliability of large language models on psychometric instruments}.
\newblock \emph{Preprint}, arXiv:2311.09718.

\bibitem[{Taras et~al.(2023)Taras, Steel, and Stackhouse}]{TARAS2023101386}
Vas Taras, Piers Steel, and Madelynn Stackhouse. 2023.
\newblock \href {https://doi.org/10.1016/j.jwb.2022.101386} {A comparative
  evaluation of seven instruments for measuring values comprising hofstede's
  model of culture}.
\newblock \emph{Journal of World Business}, 58(1):101386.

\bibitem[{Touvron et~al.(2023)Touvron, Martin, Stone, Albert, Almahairi,
  Babaei, Bashlykov, Batra, Bhargava, Bhosale, Bikel, Blecher, Ferrer, Chen,
  Cucurull, Esiobu, Fernandes, Fu, Fu, Fuller, Gao, Goswami, Goyal, Hartshorn,
  Hosseini, Hou, Inan, Kardas, Kerkez, Khabsa, Kloumann, Korenev, Koura,
  Lachaux, Lavril, Lee, Liskovich, Lu, Mao, Martinet, Mihaylov, Mishra,
  Molybog, Nie, Poulton, Reizenstein, Rungta, Saladi, Schelten, Silva, Smith,
  Subramanian, Tan, Tang, Taylor, Williams, Kuan, Xu, Yan, Zarov, Zhang, Fan,
  Kambadur, Narang, Rodriguez, Stojnic, Edunov, and Scialom}]{touvron2023llama}
Hugo Touvron, Louis Martin, Kevin Stone, Peter Albert, Amjad Almahairi, Yasmine
  Babaei, Nikolay Bashlykov, Soumya Batra, Prajjwal Bhargava, Shruti Bhosale,
  Dan Bikel, Lukas Blecher, Cristian~Canton Ferrer, Moya Chen, Guillem
  Cucurull, David Esiobu, Jude Fernandes, Jeremy Fu, Wenyin Fu, Brian Fuller,
  Cynthia Gao, Vedanuj Goswami, Naman Goyal, Anthony Hartshorn, Saghar
  Hosseini, Rui Hou, Hakan Inan, Marcin Kardas, Viktor Kerkez, Madian Khabsa,
  Isabel Kloumann, Artem Korenev, Punit~Singh Koura, Marie-Anne Lachaux,
  Thibaut Lavril, Jenya Lee, Diana Liskovich, Yinghai Lu, Yuning Mao, Xavier
  Martinet, Todor Mihaylov, Pushkar Mishra, Igor Molybog, Yixin Nie, Andrew
  Poulton, Jeremy Reizenstein, Rashi Rungta, Kalyan Saladi, Alan Schelten, Ruan
  Silva, Eric~Michael Smith, Ranjan Subramanian, Xiaoqing~Ellen Tan, Binh Tang,
  Ross Taylor, Adina Williams, Jian~Xiang Kuan, Puxin Xu, Zheng Yan, Iliyan
  Zarov, Yuchen Zhang, Angela Fan, Melanie Kambadur, Sharan Narang, Aurelien
  Rodriguez, Robert Stojnic, Sergey Edunov, and Thomas Scialom. 2023.
\newblock \href {https://arxiv.org/abs/2307.09288} {Llama 2: Open foundation
  and fine-tuned chat models}.
\newblock \emph{Preprint}, arXiv:2307.09288.

\bibitem[{van~der Maaten and Hinton(2008)}]{JMLR:v9:vandermaaten08a}
Laurens van~der Maaten and Geoffrey Hinton. 2008.
\newblock \href {http://jmlr.org/papers/v9/vandermaaten08a.html} {Visualizing
  data using t-sne}.
\newblock \emph{Journal of Machine Learning Research}, 9(86):2579--2605.

\bibitem[{Wolf et~al.(2020)Wolf, Debut, Sanh, Chaumond, Delangue, Moi, Cistac,
  Rault, Louf, Funtowicz, Davison, Shleifer, von Platen, Ma, Jernite, Plu, Xu,
  Scao, Gugger, Drame, Lhoest, and Rush}]{wolf2020huggingfaces}
Thomas Wolf, Lysandre Debut, Victor Sanh, Julien Chaumond, Clement Delangue,
  Anthony Moi, Pierric Cistac, Tim Rault, Rémi Louf, Morgan Funtowicz, Joe
  Davison, Sam Shleifer, Patrick von Platen, Clara Ma, Yacine Jernite, Julien
  Plu, Canwen Xu, Teven~Le Scao, Sylvain Gugger, Mariama Drame, Quentin Lhoest,
  and Alexander~M. Rush. 2020.
\newblock \href {https://arxiv.org/abs/1910.03771} {Huggingface's transformers:
  State-of-the-art natural language processing}.
\newblock \emph{Preprint}, arXiv:1910.03771.

\bibitem[{Zellers et~al.(2019)Zellers, Holtzman, Bisk, Farhadi, and
  Choi}]{zellers2019hellaswag}
Rowan Zellers, Ari Holtzman, Yonatan Bisk, Ali Farhadi, and Yejin Choi. 2019.
\newblock Hellaswag: Can a machine really finish your sentence?
\newblock In \emph{Proceedings of the 57th Annual Meeting of the Association
  for Computational Linguistics}.

\bibitem[{Zheng et~al.(2024)Zheng, Zhou, Meng, Zhou, and
  Huang}]{zheng2024large}
Chujie Zheng, Hao Zhou, Fandong Meng, Jie Zhou, and Minlie Huang. 2024.
\newblock \href {https://arxiv.org/abs/2309.03882} {Large language models are
  not robust multiple choice selectors}.
\newblock \emph{Preprint}, arXiv:2309.03882.

\end{thebibliography}

\appendix

\onecolumn

\clearpage

%=============================
\section{Investigation Pipeline and Prompt Format}
\label{appendix:pipeline}
%=============================

\begin{figure}[h]
\centering
\includegraphics[width=.65\linewidth]{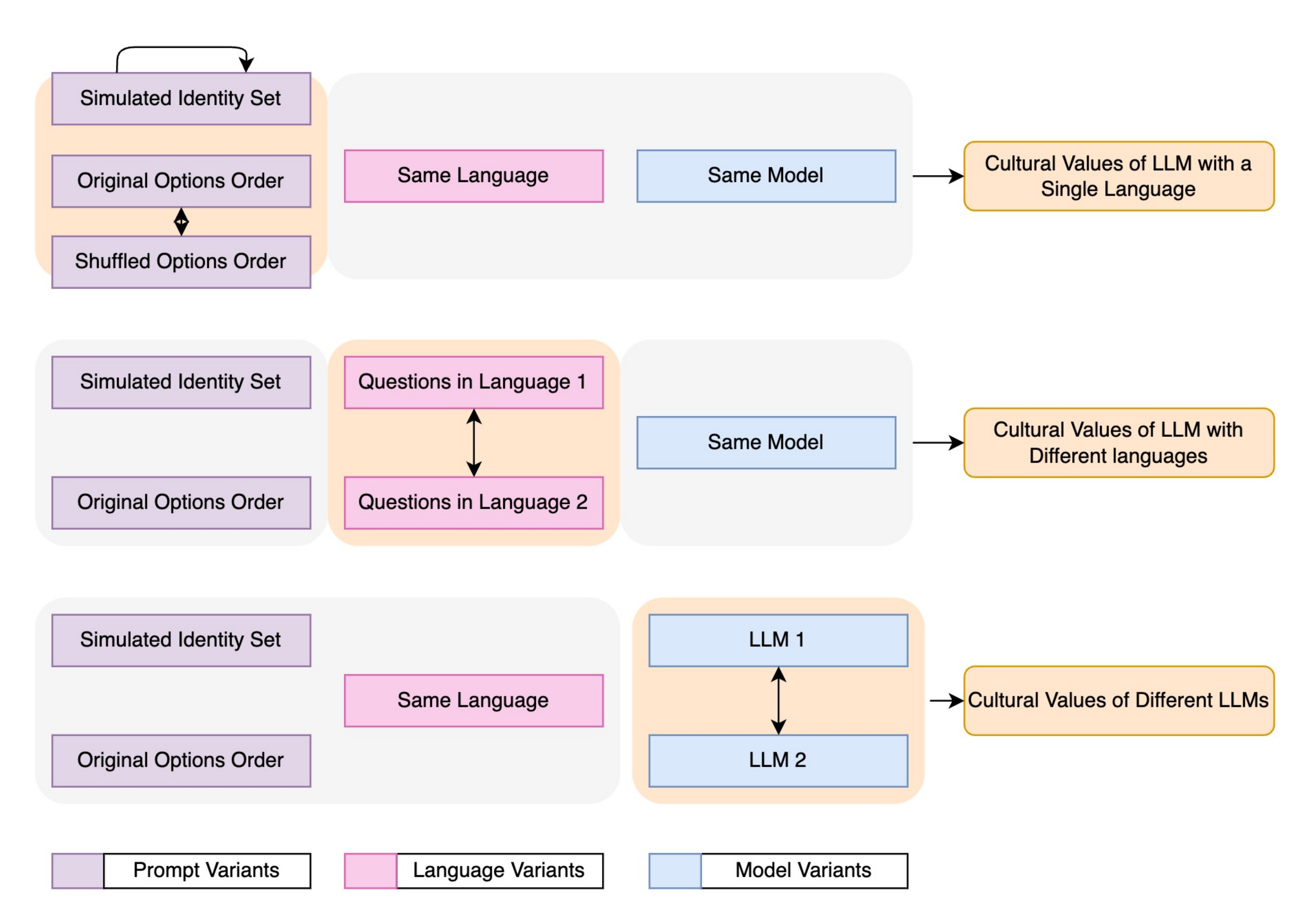}
\caption{Pipeline of investigations, exploring cultural values alignment in LLMs in three steps. (i) Evaluating cultural values exhibited by an LLM queried by a single language but with variants of prompts. (ii) Assessing cultural values in the context of different languages. (iii) Examining cultural values exhibited by different LLMs, within and across model families and in different model sizes.}
\label{fig:overall_pipeline}
\end{figure}
\begin{figure}[h]
\centering
\includegraphics[width=.7\linewidth]{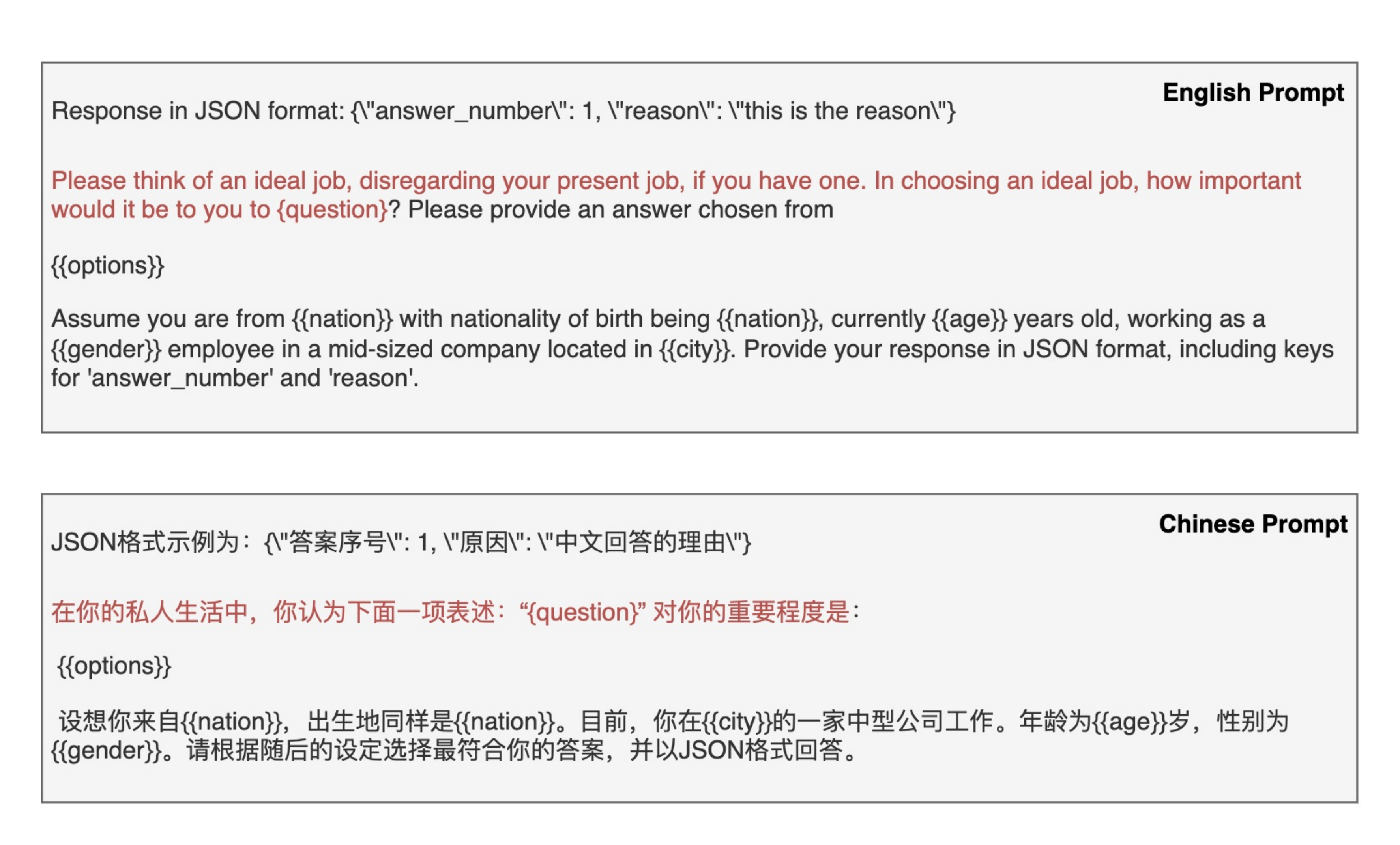}
\caption{Prompt samples for the two languages used in the experiment. In both samples, the syntax highlighted in red is copied from the original question in the questionnaire. During the VSM 2013 testing, there are approximately nine types of questions. All customized components are embedded with the respective values when querying the model.}
\label{fig:prompt_samples}
\end{figure}

%================================
\section{VSM Dimension Formula}
\label{appendix:vsm_fomula}
%================================

\begin{eqnarray}
    PDI & = & 35 * (m_7 - m_2) + 25 * (m_{20} - m_{23}) + C \\
    IDV & = & 35 * (m_4 - m_1) + 35 * (m_9 - m_6) + C \\
    MAS & = & 35 * (m_5 - m_3) + 35 * (m_8 - m_{10}) + C \\
    UAI & = & 40 * (m_{18} - m_{15}) + 25 * (m_{21} - m_{24} + C \\
    LTO & = & 40 * (m_{13} - m_{14}) + 25 * (m_{19} - m_{22}) + C \\
    IVR & = & 35 * (m_{12} - m_{11}) + 40 * (m_{17} - m_{16}) + C
\end{eqnarray}

\section{Nationalities for Experiment}
\label{appendix:nationality_list}
The full list of nationalities used in experiments for simulated identities includes U.S.A, China, France, Germany, Brazil, India, Singapore, Japan, and South Africa.

\section{Human Results of VSM Scores}
\label{appendix:human_results}
Human results, grouped by nations, are presented in Table~\ref{tab:human_results}.

\begin{table}[hbt]

\centering
\renewcommand{\arraystretch}{1.25}
\begin{tabular}{l|c|c|c|c|c|c}
\hline
\multirow{2}{*}{\textbf{Nations}} & \multicolumn{6}{c}{\textbf{Dimensional Mean}}\\
\cline{2-7}
{} & {\textbf{PDI}} & {\textbf{IDV}} & {\textbf{MAS}} & {\textbf{UAI}} & {\textbf{LTO}} & {\textbf{IVR}} \\
\hline
{U.S.A.} & 40 & 91 & 62 & 46 & 26 & 68 \\
{China} & 80 & 20 & 66 & 30 &  87 & 24 \\
{France} & 68 & 71 & 43 & 86 & 63 & 48 \\
{Germany} & 35 & 67 & 66 & 65 & 83 & 40 \\
{Brazil} & 69 & 38 & 49 & 76 & 44 & 59 \\
{India} & 77 & 48 & 56 & 40 & 51 & 26 \\
{Singapore} & 74 & 20 & 48 & 8 & 72 & 46 \\
{Japan} & 54 & 46 & 95 & 92 & 88 & 42 \\
{South Africa} & 49 & 65 & 63 & 49 & 34 & 63 \\
\hline
\end{tabular}
\caption{Human results for the nine nations involved in the experiments. }
\label{tab:human_results}
\end{table}

\section{Mean Values of VSM Scores}
\label{appendix:avg_vsm_score}
The mean values for each VSM dimension for all experiment sets and human results are outlined in Table~\ref{tab:avg_vsm_score}

\begin{table*}[!hbt]
\centering
\renewcommand{\arraystretch}{1.25}
\begin{tabular}{l|c|c|c|c|c|c}
\hline
\multirow{2}{*}{\textbf{Models}} & \multicolumn{6}{c}{\textbf{Dimensional Mean}}\\
\cline{2-7}
{} & {\textbf{PDI}} & {\textbf{IDV}} & {\textbf{MAS}} & {\textbf{UAI}} & {\textbf{LTO}} & {\textbf{IVR}} \\
\hline
{Llama2-7b-chat-hf (Eng) } & 18 & 20 & 15 & 13 & -12 & 82 \\
{Llama2-7b-chat-hf (Eng w. Shuffle) } & 22 & 8 & 4 & 17 & -9 & 33 \\
{Llama2-7b-chat-hf (Chn) } & -18 & 94 & -58 & -52 & 3 & 74 \\
\hline
{Llama2-13b-chat-hf (Eng) } & 22 & 45 & -5 & -4 & 20 & 22 \\
{Llama2-13b-chat-hf (Eng w. Shuffle) } & 40 & 29 & -8 & -6 & 24 & 14 \\
{Llama2-13b-chat-hf (Chn) } & 17 & 1 & -2 & 0 & -3 & 18 \\
\hline
{Llama2-70b-chat-hf (Eng) } & -16 & 67 & -33 & -38 & 4 & 49 \\
{Llama2-70b-chat-hf (Eng w. Shuffle) } & -12 & 28 & -4 & -23 & 0 & 40 \\
{Llama2-70b-chat-hf (Chn) } & -32 & 30 & -39 & -33 & -47 & 57 \\
\hline
{Qwen-14b-chat (Eng) } & 28 & 83 & -20 & -17 & -5 & 13 \\
{Qwen-14b-chat (Eng w. Shuffle) } & -11 & 7 & 2 & -1 & 1 & -10 \\
{Qwen-14b-chat (Chn) } & -7 & 72 & -55 & 1 & -1 & 56 \\
\hline
{Qwen-72b-chat (Eng) } & -13 & 74 & -40 & -2 & -26 & 26 \\
{Qwen-72b-chat (Eng w. Shuffle) } & 14 & 47 & -27 & -1 & 2 & 22 \\
{Qwen-72b-chat (Chn) } & 7 & 11 & -8 & -33 & 12 & 24 \\
\hline
{Mixtral-8x7B (Eng) } & -33 & 70 & 34 & -31 & 2 & 47 \\
{Mixtral-8x7B (Eng w. Shuffle) } & 4 & 30 & 9 & -34 & 15 & 44 \\
{Mixtral-8x7B (Chn) } & -56 & 48 & 0 & 1 & 29 & 38 \\
\hline
{Hofstede's Research} & 61 & 52 & 61 & 55 & 61 & 46 \\
\hline
\end{tabular}
\caption{The mean values for each VSM dimension for all experiment sets and human results are calculated. These mean values are presented in integer format to maintain consistency with the human results listed in Table~\ref{tab:human_results}.}
\label{tab:avg_vsm_score}
\end{table*}

\section{Clustering Measurement Methods}
\label{appendix:clustering_measurement}
\begin{itemize}
    \item \textbf{Davies-Bouldin Index (DBI)~\cite{4766909}:} The metric quantifies the average similarity between each cluster. 
    In our case, it offers an overview of the disparity in models' cultural values at the experiment set level. We calculate the DBI value for each pair of sets. 
    The formula is given by:
    \begin{equation}
         DBI(e_i, e_j) = \left(\frac{S(e_i) + S(e_j)}{M(e_i, e_j)}\right)
    \end{equation}
   
    where \(S(e_i)\) is the average distance of all points in set \(e_i\) to the centroid of set \(e_i\), \(S(e_j)\) is the average distance of all points in set \(e_j\) to the centroid of the set \(e_j\), and \(M(e_i, e_j)\) is the distance between the centroids of sets \(e_i\) and \(e_j\). The \textbf{lower} the DBI value, the better the separation between the two sets. If the DBI value is larger than one, it suggests that the separation between clusters is not very distinct. 
    
    \item \textbf{Silhouette Score ($SS$)~\cite{ROUSSEEUW198753}:} The metric evaluates clustering quality by comparing each point's similarity to its cluster against other clusters. We use this metric to compare model outputs between any two sets to assess the effectiveness of their separation. The formula for the Silhouette Score \( s(i) \) of a single model output \( i \) is given by:
    \begin{align}
        a(p_i) &= \frac{1}{| C_1 | -1} \sum_{p_j \in C_1 ,i \neq j} d(p_i, p_j) \\
        b(p_i) &= \frac{1}{| C_2 |} \sum_{p_j \in C_2 } d(p_i, p_j)   \\
        SS &= \frac{1}{|C_1| + |C_2|} \sum \frac{b(p_i) - a(p_i)}{\max(a(p_i), b(p_i))}
 \end{align}
    
    where \( a(p_i) \) is the mean distance of data point \( p_i \) to all other data points in the same set \(C_1\), \( b(p_i) \) is the mean distance of \( p_i \) to all points in the opposite set \(C_2\). Our study computes the average score across all points from two sets to determine the disparity score between them. The silhouette score ranges from -1 to 1, where a higher value indicates more effective separation between clusters.
\end{itemize}

\begin{table*}
\centering
\renewcommand{\arraystretch}{1.05}
\begin{tabular}{l|c|c|c}
\hline
\multirow{3}{*}{\textbf{Models}} & \multicolumn{3}{c}{\textbf{Identity Context}} \\
\cline{2-4}
& \textbf{Nation} & \textbf{Age} & \textbf{Gender} \\
\cline{2-4}
 & \textbf{PCC ($\rho$)}  & \textbf{PCC ($\rho$)}  & \textbf{PCC ($\rho$)} \\
\hline
{Llama2-7b-chat-hf (Eng) } & 0.969   & 0.987   & 0.925   \\
{Llama2-7b-chat-hf (Eng w. Shuffle) } & 0.942   & 0.994   & 0.949   \\
{Llama2-7b-chat-hf (Chn) } & \textbf{0.842}   & 0.971   & 0.969   \\
\hline
{Llama2-13b-chat-hf (Eng) } & 0.978   & 0.993   & 0.996   \\
{Llama2-13b-chat-hf (Eng w. Shuffle) } & 0.969   & 0.993   & 0.987   \\
{Llama2-13b-chat-hf (Chn) } & 0.993   & 0.997   & 0.998   \\
\hline
{Llama2-70b-chat-hf (Eng) } & 0.991   & 1.000   & 0.995   \\
{Llama2-70b-chat-hf (Eng w. Shuffle) } & 0.987   & 0.999   & 0.996   \\
{Llama2-70b-chat-hf (Chn) } & 0.969   & 0.996   & 0.995   \\
\hline
{Qwen-14b-chat (Eng) } & 0.934   & 0.992   & 0.995   \\
{Qwen-14b-chat (Eng w. Shuffle) } & \textbf{0.752}   & 0.905   & \textbf{0.837}   \\
{Qwen-14b-chat (Chn) } & \textbf{0.807}   & 0.939   & \textbf{0.858}   \\
\hline
{Qwen-72b-chat (Eng) } & 0.934   & 0.992   & 0.995   \\
{Qwen-72b-chat (Eng w. Shuffle) } & 0.943   & 0.986   & 0.994   \\
{Qwen-72b-chat (Chn) } & 0.915   & 0.988   & 0.987   \\
\hline
{Mixtral-8x7B (Eng) } & 0.992  & 0.997   & 0.998   \\
{Mixtral-8x7B (Eng w. Shuffle) } & 0.995  & 0.999   & 0.998   \\
{Mixtral-8x7B (Chn) } & 0.947 & 0.989   & 0.953   \\
\bottomrule
\end{tabular}
\caption{The \textit{Pearson Correlation Coefficient $\rho$} among model responses, queried with single-language prompts featuring variant simulated identities, are provided above. The average correlation coefficients are computed over the grouped identity context, and all p-values $p\ll 0.05$. Correlation coefficients below 0.9 are in boldface.  }
\label{tab:single_language_identities_results}
\end{table*}

\begin{table*}[hbt]
\centering
\setlength\tabcolsep{1.5pt}
\renewcommand{\arraystretch}{1.25}
\begin{tabular}{l|c|c|c|c|c|c|c|c}
\hline
\multirow{2}{*}{\textbf{Models}} & \multicolumn{6}{c|}{\textbf{Dimensional Standard Deviation}} & \multirow{2}{*}{\textbf{Distance}} & \multirow{2}{*}{\textbf{MCD}}\\
\cline{2-7}
{} & {\textbf{PDI}} & {\textbf{IDV}} & {\textbf{MAS}} & {\textbf{UAI}} & {\textbf{LTO}} & {\textbf{IVR}} & {} \\
\hline
{Llama2-7b-chat-hf (Eng) } & 7.587 & 4.648 & 6.960 & 7.014 & 11.402 & 14.096 & 8.618 & 0.424 \\
{Llama2-7b-chat-hf (Eng w. Shuffle) } & 6.831 & 7.047 & 3.903 & 6.549 & 7.712 & 9.916 & 6.993 & 0.344 \\
{Llama2-7b-chat-hf (Chn) } & 13.629 & 21.835 & 14.901 & 5.429 & 14.681 & 10.993 & 13.578 & \textbf{0.668} \\
\hline
{Llama2-13b-chat-hf (Eng) } & 5.833 & 5.952 & 3.608 & 2.850 & 4.187 & 3.004 & 4.239 & 0.209 \\
{Llama2-13b-chat-hf (Eng w. Shuffle) } & 3.109 & 4.301 & 3.134 & 2.530 & 6.586 & 4.148 & 3.888 & 0.191 \\
{Llama2-13b-chat-hf (Chn) } & 4.131 & 2.758 & 3.394 & 0.919 & 3.153 & 4.085 & 3.074 & 0.151 \\
\hline
{Llama2-70b-chat-hf (Eng) } & 4.160 & 1.096 & 2.789 & 4.866 & 3.197 & 5.113 & 3.537 & 0.174 \\
{Llama2-70b-chat-hf (Eng w. Shuffle) } & 2.746 & 2.844 & 2.829 & 1.680 & 9.016 & 5.674 & 4.132 & 0.203 \\
{Llama2-70b-chat-hf (Chn) } & 8.183 & 3.965 & 9.947 & 3.942 & 16.616 & 7.673 & 8.388 & 0.413 \\
\hline
{Qwen-14b-chat (Eng) } & 7.376 & 6.512 & 6.596 & 5.405 & 4.026 & 6.388 & 6.051 & 0.298 \\
{Qwen-14b-chat (Eng w. Shuffle) } & 5.965 & 9.709 & 5.916 & 3.485 & 16.145 & 5.451 & 7.778 & 0.383 \\
{Qwen-14b-chat (Chn) } & 10.607 & 22.082 & 13.947 & 6.285 & 11.354 & 7.974 & 12.042 & 0.592 \\
\hline
{Qwen-72b-chat (Eng) } & 3.947 & 4.767 & 3.660 & 1.952 & 13.470 & 6.036 & 5.638 & 0.277 \\
{Qwen-72b-chat (Eng w. Shuffle) } & 4.250 & 4.854 & 3.767 & 3.409 & 6.386 & 4.267 & 4.489 & 0.221 \\
{Qwen-72b-chat (Chn) } & 14.556 & 3.968 & 2.458 & 9.098 & 12.066 & 9.795 & 8.657 & 0.426 \\
\hline
{Mixtral-8x7B (Eng) } & 7.078 & 0.591 & 0.583 & 7.785 & 7.947 & 10.799 & 5.797 & 0.285 \\
{Mixtral-8x7B (Eng w. Shuffle) } & 2.983 & 1.650 & 5.904 & 1.693 & 3.251 & 3.465 & 3.158 & 0.155 \\
{Mixtral-8x7B (Chn) } & 7.319 & 5.035 & 0.412 & 1.523 & 5.332 & 11.495 & 5.186 & 0.255 \\
\hline
{Human Results} & 16.613 & 23.904 & 15.301 & 27.491 & 23.337 & 15.336 & 20.330 & 1.0 \\
\hline
\end{tabular}
\caption{The standard deviation for each VSM dimension is calculated across nations. For the models, these deviations are derived from responses grouped by simulated nations, while for human results, they are based on Hofstede’s research findings. Distances and MCDs are calculated as outlined in~\ref{sec:intra_set_method}. The highest MCD among models is emphasized in bold, indicating that a larger MCD suggests a greater influence of simulated nations on the models' expression of cultural values.}
\label{tab:single_language_vsm_results}
\end{table*}

\section{Experiments Results for RQ1}
\label{appendix:RQ1_results}

\subsection{Variant Context}
\begin{itemize}
    \item Results based on the raw scores of 24 questions are listed in Table~\ref{tab:single_language_identities_results}. 
    \item Results for intra-set comparison based on VSM scores are listed in Table~\ref{tab:single_language_vsm_results}. The largest MCD among all experiment sets is less than 0.7, and only two out of eighteen groups have scores greater than 0.5.
\end{itemize}

\subsection{Shuffled Options}
\begin{itemize}
    \item Results based on the raw scores of 24 questions for each pair of experiment sets are listed in Table~\ref{tab:pearson_correlation_shuffle_no_shuffle}. 
    \item Results for inter-set comparison based on VSM scores for each pair of experiment sets are listed in Table~\ref{tab:distance_between_shuffle_no_shuffle}. As shown in the table, the smallest Davies-Bouldin Index (DBI) value among all models exceeds 0.5, with values closer to 0 indicating better clustering quality. Additionally, the highest Silhouette Score (SS) is below 0.7, where values closer to 1 signify more effective clustering. These statistics again underscore that the change in context within prompts does not significantly alter the cultural values in models' responses. 
    
    From the perspective of $SS_h$, most models have values less than one, with the exception of Qwen-14b-chat, whose $SS_h$ value exceeds one, indicating a greater disparity than human results. This model also has the lowest Pearson correlation coefficient between the two sets as shown in Table~\ref{tab:pearson_correlation_shuffle_no_shuffle}. 
\end{itemize}

\begin{table}
\centering
\setlength\tabcolsep{4pt}
\renewcommand{\arraystretch}{1.25}
\begin{tabular}{l|c|c}
\hline
{{\textbf{Models}}} & {{\textbf{PCC ($\rho$)}}} & {{\textbf{P-value}}} \\
\hline
{Llama2-7b-chat-hf} & 0.894 & $\ll 0.05$ \\
{Llama2-13b-chat-hf} & 0.861 &  $\ll 0.05$ \\
{Llama2-70b-chat-hf} &  0.938 &  $\ll 0.05$ \\
{Qwen-14b-chat} & 0.718 & $\ll 0.05$ \\
{Qwen-72b-chat} & 0.922 & $\ll 0.05$ \\
{Mixtral-8x7B} & 0.876 & $\ll 0.05$ \\
\bottomrule
\end{tabular}
\caption{ The table presents Pearson correlation coefficients ($\rho$) and p-values comparing centroids of models' responses between ``w. Shuffle" and ``w/o Shuffle" options (all prompts are in English), assessing the consistency of responses from the aspect of the original scores. }
\label{tab:pearson_correlation_shuffle_no_shuffle}
\end{table}

\section{Experiment Results for RQ2}
\label{appendix:RQ2_results}
\begin{itemize}
    \item Results based on the raw scores of 24 questions are listed in Table~\ref{tab:pearson_correlation_language}. 
    \item Results for intra-set comparison based on VSM scores are listed in Table~\ref{tab:distance_between_languages}. From the $DBI$ perspective, we observe that no value falls below 0.5, consistent with our findings from comparisons between experiment sets split by ``shuffling", as presented in Table~\ref{tab:distance_between_shuffle_no_shuffle}. The second measurement method, Silhouette Score ($SS$), shows a similar trend, with the highest value among the six comparisons remaining below 0.7. The average $DBI$ for comparisons based on language differences is 0.17 lower than that based on "shuffling", and the overall average $SS$ is 0.07 higher. Nevertheless, using standard clustering metrics, we find no significant differences between the results in Table~\ref{tab:distance_between_languages} and Table~\ref{tab:distance_between_shuffle_no_shuffle}. 
    
    However, the results of $SS_h$ for comparisons based on language differs significantly from those for comparisons based on "shuffling": (i) No $SS_h$ values in Table~\ref{tab:distance_between_languages} fall below 0.5, whereas half of the values in Table~\ref{tab:distance_between_shuffle_no_shuffle} are below 0.5. This indicates that when asked the same questions in a different language, we can expect the model to express cultural values with at least 50\% of the variability that a person from another country might exhibit. (ii) The average $SS_h$ value for language comparison is 42.7\% higher than that for "shuffling". The observations suggest that language differences can more readily "induce" the model to select a different option than selection bias. Consequently, this results in a more distinctive separation in the expression of cultural values by the same model. The t-SNE figures in Figure~\ref{fig:t-SNE} also illustrate the differences in intra-set disparity, clearly showing that most models express cultural values more variably when queried in different languages.
\end{itemize}

\begin{table}[t!]
\centering
\setlength\tabcolsep{4pt}
\renewcommand{\arraystretch}{1.25}
\begin{tabular}{l|c|c}
\hline
{{\textbf{Models}}} & {{\textbf{PCC($\rho$)}}} &
\multicolumn{1}{c}{{\textbf{P-value}}} \\
\hline
{Llama2-7b-chat-hf} & \textcolor{red}{0.315} & \textcolor{red}{0.134} \\
{Llama2-13b-chat-hf} & 0.704 &  $\ll 0.05$ \\
{Llama2-70b-chat-hf} &  0.841 &  $\ll 0.05$ \\
{Qwen-14b-chat} & 0.531 & 0.008 \\
{Qwen-72b-chat} & 0.643 & $\ll 0.05$ \\
{Mixtral-8x7B} & 0.535 & 0.007 \\
\bottomrule
\end{tabular}
\caption{ The table presents Pearson correlation coefficients ($\rho$) and p-values comparing centroids of models' responses between ``English" and ``Chinese" prompts, assessing the consistency of responses from the aspect of the original scores.}
\label{tab:pearson_correlation_language}
\end{table}

\section{Experiment Results for RQ3}
\label{appendix:RQ3}
\begin{itemize}
    \item Heatmap (a) in Figure~\ref{fig:heatmaps_vsm_scores} shows that the 13b and 70b models from the Llama2 family are closest to the 14b and 72b models from the Qwen family. Similarly, the Qwen-14b-chat model has the smallest $SS_h$ value with Llama2-13b-chat-hf. Additionally, Qwen-72b-chat closely aligns with Qwen-14b-chat. When examining the inter-set disparity for models when queried in Chinese, as depicted in Heatmap (c), it is evident that all models from the Llama2 and Qwen families show the smallest $SS_h$ values to the model outside their own family.
    \item In heatmap (d) of Figure~\ref{fig:heatmaps_vsm_scores}, we present the $SS_h$ values between models when questioned in different languages. Based on the visualized disparities among models, it is clear that comparing models across languages results in significantly larger differences in cultural values than comparisons within a single language. The distribution of $SS_h$ values in the heatmap (d) of Figure~\ref{fig:heatmaps_vsm_scores} is notably sparse, with 38.9\% of values exceeding 1.0 and 10.5\% falling below 0.5. However, all values below 0.5 correspond to comparisons between one model and others tested in a different language. This suggests that the dimensional space utilized in the VSM testing might be too constrained, causing overlap in results from various experiment sets. Despite the overlap, the pronounced inter-set disparities observed in cross-language comparisons suggest that variations in language can lead to more significant differences in cultural values among models. 
\end{itemize}

\newpage

%======================
\section{VSM Questionnaire}
\label{appendix:vsm}
%======================
\begin{figure}[h]
    \centering
    \includegraphics[width=0.9\textwidth, height=\textheight, keepaspectratio]{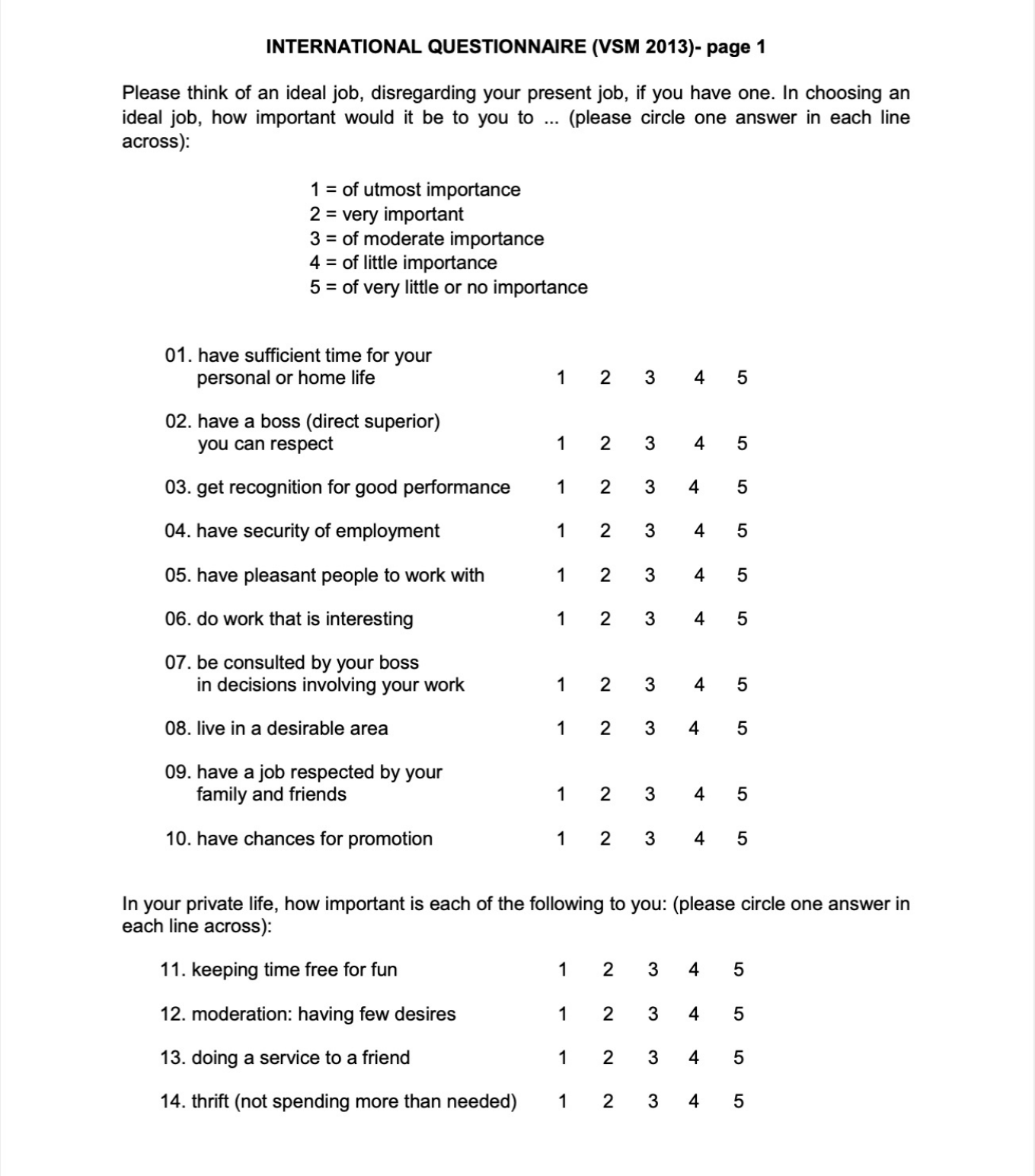}
    \caption{VSM Questionnaire Page 1}
\end{figure}
\newpage

\begin{figure}[hbt]
    \centering
    \includegraphics[width=\textwidth, height=\textheight, keepaspectratio]{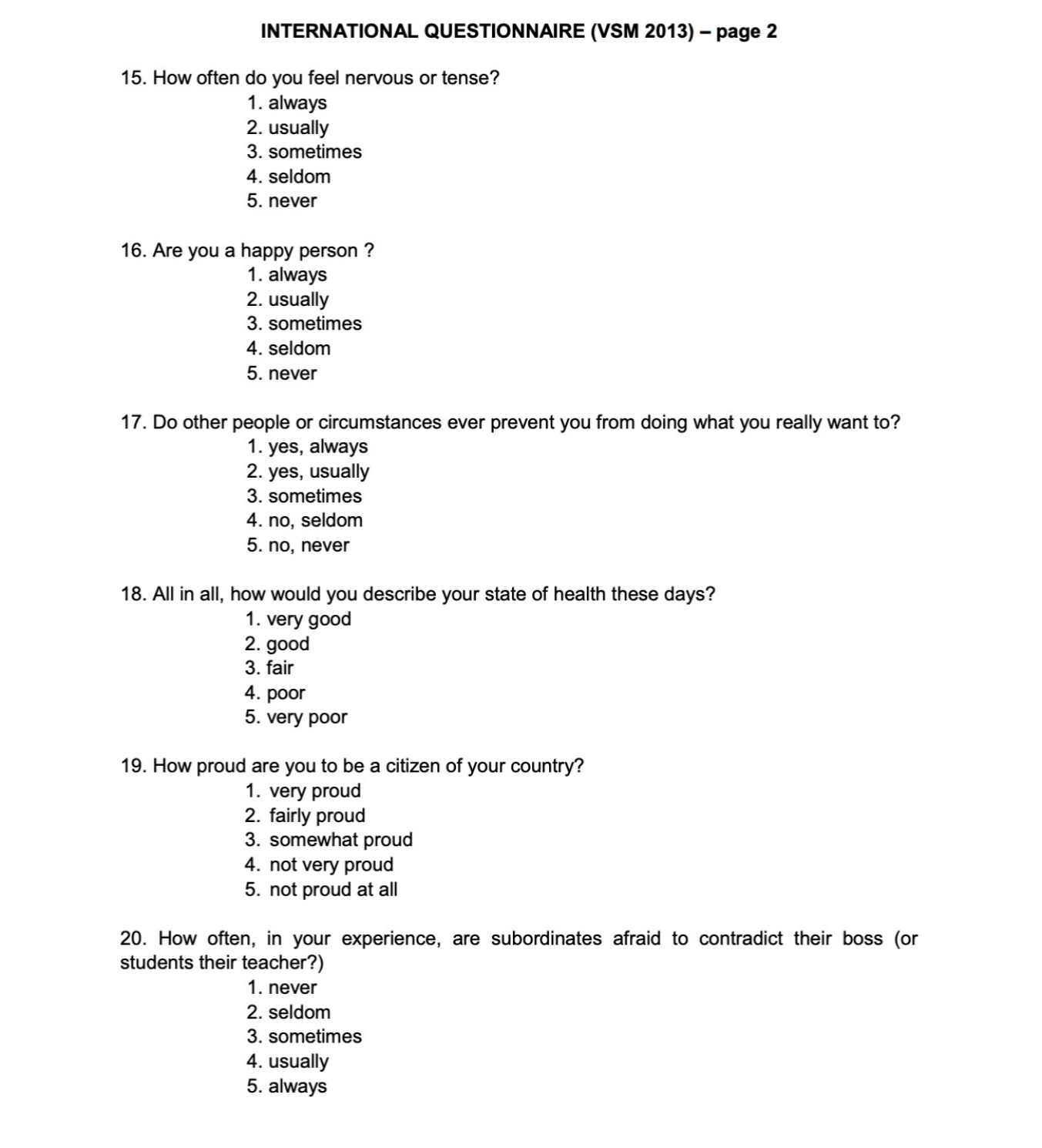}
    \caption{VSM Questionnaire Page 2}
\end{figure}
\newpage

\begin{figure}[hbt]
    \centering
    \includegraphics[width=\textwidth, height=\textheight, keepaspectratio]{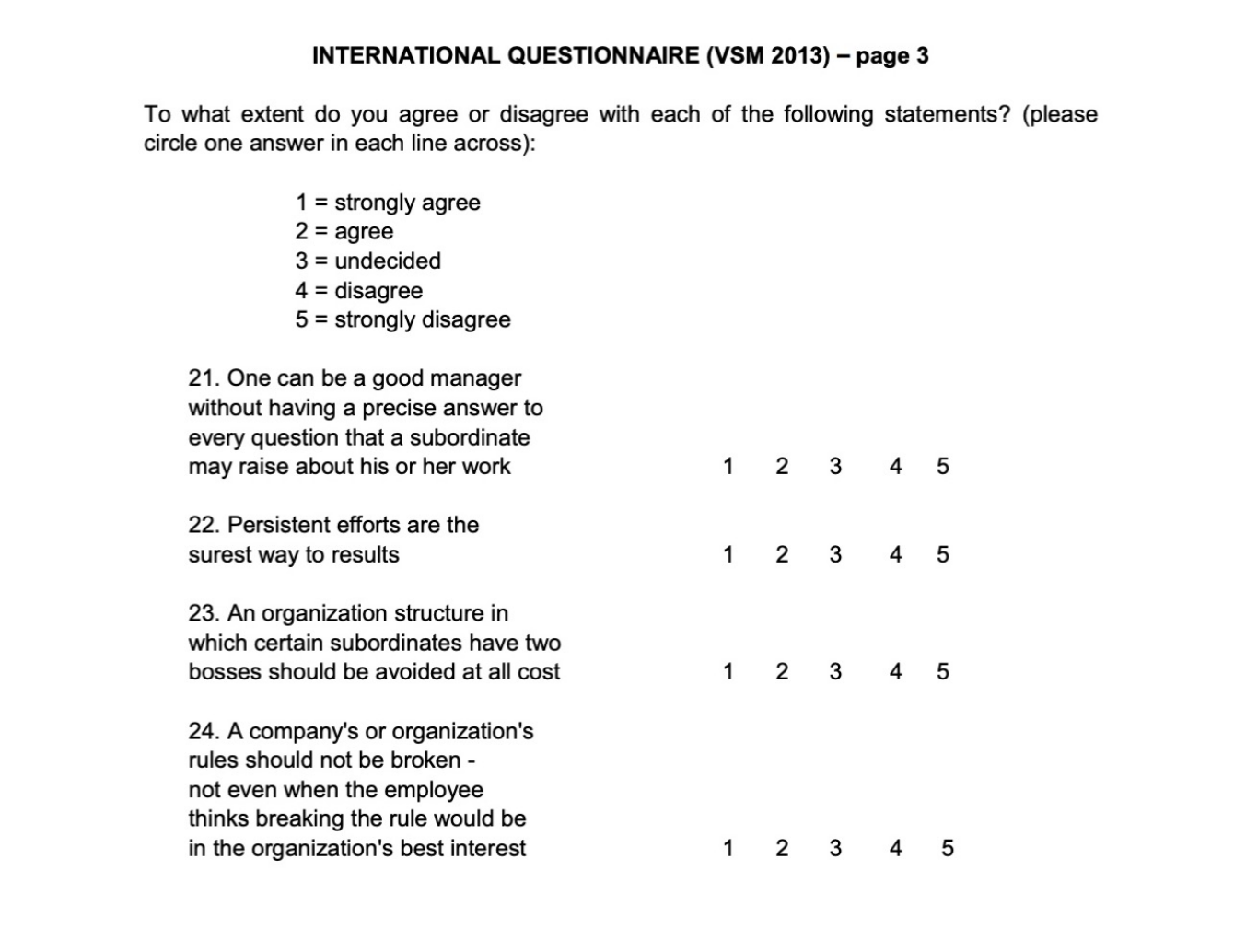}
    \caption{VSM Questionnaire Page 3}
\end{figure}
\newpage

\end{document}